\documentclass[a4paper,12pt]{article}
\usepackage{epsfig,psfig,epsf}

\usepackage{citesort}
 \normalsize

\newcommand{\bea}{\begin{eqnarray}}
\newcommand{\eea}{\end{eqnarray}}

\begin{document}

\def\lsim{\raise0.3ex\hbox{$\;<$\kern-0.75em\raise-1.1ex\hbox{$\sim\;$}}} 

\def\gsim{\raise0.3ex\hbox{$\;>$\kern-0.75em\raise-1.1ex\hbox{$\sim\;$}}}

\def\Frac#1#2{\frac{\displaystyle{#1}}{\displaystyle{#2}}}
\def\no{\nonumber\\}

\def\dobox#1#2{\centerline{\epsfxsize=#1\epsfig{file=#2, width=10cm,
height=5.5cm, angle=0}}}

\def\dofigure#1#2{\centerline{\epsfxsize=#1\epsfig{file=#2, width=8cm, 
height=10cm, angle=0}}}
%
\def\dofig#1#2{\centerline{\epsfxsize=#1\epsfig{file=#2, width=7cm, 
height=10cm, angle=-90}}}
\def\dofigspec#1#2{\centerline{\epsfxsize=#1\epsfig{file=#2, width=7cm, 
height=10cm, angle=-90}}}
\def\dofigA#1#2{\centerline{\epsfxsize=#1\epsfig{file=#2, width=12cm, 
height=3cm, angle=0}\vspace{0cm}}}
\def\dofigs#1#2#3{\centerline{\epsfxsize=#1\epsfig{file=#2, width=6cm, 
height=7.5cm, angle=-90}\hspace{0cm}
\hfil\epsfxsize=#1\epsfig{file=#3,  width=6cm, height=7.5cm, angle=-90}}}
\def\dotwofigs#1#2#3{\centerline{\epsfxsize=#1\epsfig{file=#2, width=8cm, 
height=6cm, angle=0}\hspace{0cm}
\hfil\epsfxsize=#1\epsfig{file=#3,  width=8cm, height=8cm, angle=0}}}
\def\dofourfigs#1#2#3#4#5{\centerline{
\epsfxsize=#1\epsfig{file=#2, width=6cm,height=7.5cm, angle=-90}
\hspace{0cm}
\hfil
\epsfxsize=#1\epsfig{file=#3,  width=6cm, height=7.5cm, angle=-90}}

\vspace{0.5cm}
\centerline{
\epsfxsize=#1\epsfig{file=#4, width=6cm,height=7.5cm, angle=-90}
\hspace{0cm}
\hfil
\epsfxsize=#1\epsfig{file=#5,  width=6cm, height=7.5cm, angle=-90}}
}

\def\dosixfigs#1#2#3#4#5#6#7{\centerline{
\epsfxsize=#1\epsfig{file=#2, width=6cm,height=7cm, angle=-90}
\hspace{-1cm}
\hfil
\epsfxsize=#1\epsfig{file=#3,  width=6cm, height=7cm, angle=-90}}

\centerline{
\epsfxsize=#1\epsfig{file=#4, width=6cm,height=7cm, angle=-90}
\hspace{-1cm}
\hfil
\epsfxsize=#1\epsfig{file=#5,  width=6cm, height=7cm, angle=-90}}

\centerline{
\epsfxsize=#1\epsfig{file=#6, width=6cm,height=7cm, angle=-90}
\hspace{-1cm}
\hfil
\epsfxsize=#1\epsfig{file=#7,  width=6cm, height=7cm, angle=-90}}
}

\def\no{\nonumber\\}
\def\slash#1{\ooalign{\hfil/\hfil\crcr$#1$}}
\def\ep{\eta^{\prime}}

%
\begin{titlepage}
\vspace*{-2cm}
\begin{flushright}
HIP-2006-20/TH
\end{flushright}

{\Large
\begin{center}
{\bf Photon propagation in magnetic and electric fields with
scalar/pseudoscalar couplings: a new look}
\end{center}
}
\vspace{.5cm}

\begin{center}
{Emidio Gabrielli$^{a}$, }
{Katri Huitu$^{a,b}$, }
{Sourov Roy$^{a}$ }
\\[5mm]
{$^{a}$\textit{Helsinki Institute of Physics,
P.O.B. 64, 00014 University of  Helsinki, Finland }}\\[0pt]
{$^{b}$\textit{Div. of HEP, Dept. of Physical Sciences,  P.O.B. 64,
00014 University of  Helsinki, Finland }}\\
[10pt]
\vspace{1.5cm}
\begin{abstract}
We consider the minimal coupling of two photons to neutral scalar 
and pseudoscalar fields, as for instance in the case 
of the Higgs boson and axion, respectively.
In this framework, we analyze the photon dispersion relations
in the presence of static and homogeneous external magnetic and 
electric fields, by taking into account the contribution 
of the imaginary part of the
scalar/pseudoscalar self-energy. We show that this contribution cannot be
neglected when it is of the same order as the
photon-scalar/pseudoscalar mixing term.
In addition to the usual light-like photon propagation
mode, with a refraction index $n > 1$, 
a {\it massive} mode with mass of the order of the coupled boson mass
can be induced, provided that
the external field is above a particular critical value.
Depending on the values of the external field, photon energy, and
mass of the scalar/pseudoscalar particle, the  
scalar/pseudoscalar width could induce a sizeable 
rate of photon splitting in two photons due to a strong resonant phenomenon.
This effect has no practical laboratory 
applications for the Higgs physics due to the very large critical 
external magnetic or electric fields involved, 
for a photon energy of the order of a TeV. However, it can have relevant 
consequences in the axion physics or in any other scenario
where light neutral scalar/pseudoscalar fields have minimal 
coupling with two photons.
\end{abstract}

\end{center}
\end{titlepage}

\section{Introduction}
Scalar and pseudoscalar neutral particles can have effective gauge invariant
coupling with the electromagnetic field $A_{\mu}$ due to the
higher dimensional operators.
In particular, in the case of a scalar ($\varphi_S$) 
and pseudoscalar ($\varphi_P$) fields, 
the minimal interaction Lagrangians, proportional to 
the lowest dimensional gauge invariant operators,  are given respectively by 
\bea
L_{S}&=& -\frac{1}{4\Lambda_S}\, F_{\mu\nu} F^{\mu\nu}\, \varphi_S\, ,
\label{LeffS}
\\
L_{P}&=& -\frac{1}{4\Lambda}_P\, F_{\mu\nu} \tilde{F}^{\mu\nu}\, \varphi_P\, ,
\label{LeffP}
\eea
where 
$F_{\mu\nu}=\partial_{\mu} A_{\nu}-\partial_{\nu} A_{\mu}$ 
is the electromagnetic field strength tensor, 
$\tilde{F}_{\mu\nu}=\frac{1}{2}\epsilon_{\mu\nu\alpha\beta}\, F^{\alpha\beta}$ 
its dual,  and the mass scales $\Lambda_{S,P}$ parametrize the corresponding 
couplings.

Well known examples of such couplings to two photons
can be found, for instance, in the framework of Higgs boson 
\cite{Hrev,ggH}
and axion physics \cite{PQ,WW,raffelt_book}.
The Higgs mechanism, responsible for 
the electroweak symmetry breaking (EWSB) in the Standard Model 
(SM), predicts a heavy scalar particle, the Higgs boson,
which up to now has eluded any experimental test.
The Higgs boson mass cannot be predicted in the framework of SM, 
and it ranges from the present experimental lower limit \cite{LEP,PDG} of 
114 GeV, up to the theoretical upper bound of around $800$ GeV \cite{Hrev}.
Although the Higgs boson has no tree-level coupling with two 
photons, it acquires such a coupling at 1-loop \cite{ggH}, with 
the corresponding scale $\Lambda_S$
of the order of ${\cal O}$(TeV).

On the contrary, the axion, which arises as a natural 
solution to the strong CP problem \cite{PQ,WW}, 
is a very light pseudoscalar particle
which remains undetected due to the very weak coupling to ordinary matter.
It appears as a pseudo Nambu-Goldstone boson of a spontaneously broken 
Peccei-Quinn symmetry \cite{PQ} 
and has a mass $m_A$ expected to be of the order of 
$m_A\sim {\cal O}$(meV) 
\cite{sikivie1,axion,sikivie05,axion_astro}. Moreover, since
the corresponding 
coupling $\Lambda_P$ is of the order of $10^{10}-10^{11}$ GeV if
astrophysical constraints are taken into account \cite{axion_astro},
the decay width of the axion into two photons,
being proportional to $m_A^3/\Lambda_P^2$,
is very small, implying that the axion is almost a stable 
particle on cosmic time scales.

The fact that a scalar field has an effective coupling with two photons, 
allows the $\gamma\to \varphi_S$ conversion mechanism in 
external magnetic or electric field to work by means of the Primakoff effect
\cite{Prim}.
A simple way to look at this phenomenon is the following. By 
replacing one of the photon fields in $\varphi_S \gamma\gamma$ coupling
in Eq.(\ref{LeffS})
with an external electromagnetic (EM) background field, 
the resulting $\varphi_S \gamma$ mixing term
could be regarded as an off-diagonal term of
the scalar-photon mass matrix, providing a source for the $\gamma\to \varphi_S$
transition. 
This is not in contradiction with angular momentum conservation, 
since due to the presence of the external field, the SO(3) 
rotational invariance
for the total system radiation + background is restored.
Analogous results hold for the pseudoscalar
conversion mechanism $\gamma\to \varphi_P$ as well \cite{sikivie1,axion}.

The mechanism of photon-axion conversion in external
magnetic field has been intensively studied in the literature
\cite{sikivie1,axion,sikivie05,axion_astro}, especially in 
the context of astrophysics and cosmology \cite{axion_astro}, 
as the time evolution of the quantum system of two particle states.
Due to the very large life-time of the axion, 
the photon-axion oscillation phenomenon is possible on macroscopic distances.
However, it is worth stressing that
the photon-axion conversion, with both fields on-shell, 
is not possible on strictly homogeneous and constant magnetic fields
in all space. Indeed, due to the fact that 
the axion is a massive particle, the conservation of the energy and momentum 
at the axion-photon vertex requires inhomogeneity of the
external magnetic field in order the reaction to proceed.
Therefore, the transition amplitude comes out to be
proportional to the form factor of the external field, inducing a suppression
\cite{sikivie1,axion,sikivie05}.
When the scale $(L)$ over which the magnetic field is homogeneous,
is comparable or smaller than 
the scale of the inverse of the transfered momentum 
$\Delta P\sim m_A^2/(2 E)$, the magnetic form factor could easily be
of order ${\cal O}(1)$. This is indeed the case of 
optical photons and axion masses of the order of
meV, where $\Delta P L\ll 1$. On the contrary, when
$\Delta P L\gg 1$, a strong suppression is expected from the form factor
if the magnetic field is homogeneous. 
However, the depletion induced by the form factor can be ameliorated
by embedding the system in a plasma, and tuning the plasma 
frequency, which provides an effective mass for the photon, 
to be equal to the axion mass \cite{axion_plasma}.

A particular class of experiments \cite{axion_exp1,BFRT} 
have been using the idea 
of the photon regeneration mechanism \cite{sikivie1}
in order to detect the axion in laboratory.
After a Laser beam
passes through a magnetic field, an axion component can be generated.
The original incident Laser is then 
absorbed through a thick wall, but not the axion component.
This can be then re-converted into photons by passing through a second magnet, 
the so-called phenomenon of light shining from a wall.

However, indirect effects of the axion coupling with two photons 
could give rise to
the modification of photon dispersion relations in the presence of
an external constant magnetic field \cite{MPZ}.
In particular, these effects would induce the so called `Faraday rotation',
i.e. the rotation of the plane of polarization of a plane polarized
light passing through a magnetic field  \cite{MPZ}.
The BFRT collaboration \cite{BFRT} has also performed a 
polarization experiment along these lines, but no significant deviation
has been observed \cite{axion_exp1}.
Recently, the PVLAS experiment \cite{PVLAS},
has reported the first evidence of a rotation of the polarization plane 
of light propagating through a static magnetic field. If confirmed, 
these results might point out the presence of a very light 
pseudoscalar particle weakly coupled to two photons. 
Alternative experiments have been proposed to check these results 
\cite{ringw1,ringw2}. The phenomenon of `Faraday rotation' in a medium
has also been analyzed in the context of QED \cite{Faraday}.

In this framework, the aim of the present paper is to 
investigate the photon dispersion relations
on static and homogeneous external electric or magnetic fields, by 
using the approach of the effective photon propagator.
This last one is 
obtained after integrating out the scalar/pseudoscalar degree of freedom,
or equivalently by summing up the relevant class of 
photon self-energy diagrams as done in our paper.
Apart from the different approach, 
the new aspect with respect to previous studies mainly consists 
in the fact that we are including the contribution 
of scalar/pseudoscalar self-energy
diagrams, where the effects of both real and 
imaginary parts are taken into account. In particular,
we will show that the contribution of the imaginary part cannot be neglected.
This approach 
would allow to analyze new phenomena not considered before, such as 
the effects of the scalar/pseudoscalar decay {\it width} ($\Gamma$)
on the absorptive part of the effective photon propagator.
Nevertheless, one should also expect that the  {\it width} $\Gamma$
corrections 
could induce some effects on the rotation of the polarization 
plane of the incident photon, although we 
have not considered this aspect in the present work.

Analogous studies are well known in the framework of QED, 
where the refraction index for photon propagation and absorption coefficient 
have been determined in the case of photon propagating in strong magnetic 
fields \cite{adler,pair_creat,erber}.
In our case, this task can be easily achieved
by summing up the full series of leading Feynman diagrams 
of photon self-energy and
dispersion relations can then be derived by looking at the poles
of the effective propagator.
While the real part of the scalar/pseudoscalar self-energy 
(${\rm Re}\Sigma$)
can be re-absorbed in the renormalization of the corresponding 
mass and wave function, the ${\rm Im}\Sigma$
would induce an imaginary part in the effective photon propagator,
giving rise to a non-vanishing contribution to the
photon absorption coefficient.

Due to the presence of the external field, the vacuum polarization 
and the dispersion relations will be modified with respect
to the case without external sources. This will affect only the
component of the photon polarization vector interacting with the
external field, according to the Lagrangian in Eqs.(\ref{LeffS}), 
(\ref{LeffP}), while the other one will remain unaffected
\cite{MPZ}.
In  particular, there will be
two new propagation modes in addition to the one which is not modified.
The {\it lightest} mode, corresponding to a light-like particle
with a refraction index $n >1$, and a {\it massive} mode
corresponding to a massive particle with mass of the order of the exchanged 
scalar/pseudoscalar one.
By taking into account the effect of the scalar/pseudoscalar {\it width}
$\Gamma$, 
we will show that the {\it massive} propagation mode is a solution of the 
dispersion relations only when the size of 
$\Gamma$ would be smaller or
comparable to the effect of the mixing term, or in other words
the massive solution is allowed only when 
the external magnetic or electric field is above some 
particular critical value. 

Finally, we would like to stress that the scalar/pseudoscalar
coupling with two photons can induce a sizeable contribution 
to the photon splitting process $\gamma\to \gamma\gamma$
in external magnetic or electric fields, depending
on the magnitude of the external field and the photon energy involved.
However, although suppressed by the corresponding width of
the scalar or pseudoscalar field, 
the corresponding photon absorption coefficient
may turn out to be significant due to a strong resonant phenomenon
when the external field is in the vicinity of the critical value.

These results suggest
a new class of experiments for searching indirect effects of 
the scalar/pseudoscalar coupling to two photons, like for instance
by searching for the photon splitting $\gamma\to \gamma\gamma$
in constant magnetic fields.
Notice that this process does not have any background since 
the analogous effect in QED
is strongly suppressed for laboratory magnetic fields
and photon energies in the optical or X-ray frequency range \cite{adler}. 
Therefore, any observation of photon splitting signal
in laboratory would be a clear indication of new physics effects.

Returning to the scalar coupling with two photon,
a well-known example in the standard model is provided by the Higgs boson
\cite{ggH}.
Unfortunately, in this case 
a too large critical magnetic (or electric) external field 
would be required for photon energies of the order of TeV,
with no practical laboratory applications.
Nevertheless, we stress that sizeable effects on the photon
absorption coefficient, induced by the Higgs boson, might 
be possible in astrophysical context, 
where sources of strong magnetic fields and/or very high energy photons
can be found. For example in the core of a neutron star magnetic field
can be as large as $10^{8}$ Tesla \cite{raffelt_book}, 
although in that case plasma effects should be taken into account.

The plan of the paper is the following.
In section 2, we present the results for the effective 
photon propagator in the presence of a static and homogeneous
external magnetic field.
In section 3, we provide the solutions to the photon 
dispersion relations, while
in section 4 we will show the results of the 
photon absorption coefficient.
In section 5 the phenomenological 
implications of these results will be analyzed 
and numerical results will be provided.
Finally, in section 6 we will present our conclusions.
In appendix A we report 
the exact solutions of the dispersion relations, while in appendix B we 
provide the calculation of the imaginary part of scalar(pseudoscalar) 
self-energy.

\section{Effective photon propagator}

Let us consider first the case of a scalar field and
an external constant and homogeneous magnetic field $\vec{\rm B}$ assumed
to be extended in all space.
The Lagrangian containing the pure radiation field and other 
dynamical fields, can be
obtained by decomposing the electromagnetic field strength 
$F_{\mu\nu}$ in two parts
\bea
F_{\mu\nu}=F_{\mu\nu}(A)+F_{\mu\nu}^{\rm (ext)}
\label{shift}
\eea
where 
$F_{\mu\nu}(A)=\partial_{\mu} A_{\nu} 
-\partial_{\nu} A_{\mu}$ contains the usual photon radiation field 
$A_{\mu}$, while
the $F_{\mu\nu}^{\rm (ext)}$ includes the corresponding
term induced by the external magnetic field.
After the shift in Eq.(\ref{shift}), 
the relevant Lagrangian would be given by
\bea
L=-\frac{1}{4}\, F_{\mu\nu}(A)F^{\mu\nu}(A)+
\frac{1}{2}\partial_{\mu}\varphi_S \partial^{\mu}\varphi_S -
\frac{1}{2}m^2 \varphi_S^2 -\frac{1}{4\Lambda_S} 
F_{\mu\nu}(A) F^{\mu\nu}(A) \varphi_S + L_{\rm ext}({\rm \vec{B}})
\label{L0}
\eea
where $m$ is the mass of the scalar field and 
$L_{\rm ext}({\rm \vec{B}})$ is given by
\bea
L_{\rm ext}({\rm \vec{B}})&=&-\frac{1}{4}
F_{\mu\nu}^{\rm (ext)}F^{{\rm (ext)} \mu\nu }
-\frac{1}{2}
F_{\mu\nu}^{\rm (ext)}F^{\mu\nu }(A)
-\frac{1}{2\Lambda_S} 
F^{\mu\nu}(A) F_{\mu\nu}^{\rm (ext)} \varphi_S
\nonumber\\
&-&\frac{1}{4\Lambda_S} 
F_{\mu\nu}^{\rm (ext)} F^{{\rm (ext)} \mu\nu}  \varphi_S \, .
\label{Lmixed}
\eea
Notice that the last term in Eq.(\ref{Lmixed}), 
in the case in which the external field
is constant, is a tadpole term for the scalar field $\varphi_S$. 
It is easy to see that this 
tadpole has no physical effect since it can be eliminated
by making the following constant shift 
$\varphi_S\to \varphi_S+\delta$ where $\delta=-F_{\mu\nu}^{\rm (ext)}
F^{({\rm ext}) \mu\nu}/(4\Lambda_S m^2)$.
The extra term $ -\frac{\delta}{4\Lambda_S} F_{\mu\nu}(A) F^{\mu\nu}(A)$
can be then re-absorbed in the photon wave function ($A_{\mu}$)
renormalization $A_{\mu}\to Z^{1/2}\,A_{\mu}$ with 
$Z=\left(1+\frac{\delta}{\Lambda_S}\right)^{-1}$. There
are no other effects of the shifting, apart from adding an extra constant
term to the Lagrangian.\footnote{Regarding the 
mixed term $F_{\mu\nu}^{({\rm ext})} F^{\mu\nu }(A)$, we will not
consider it in the analysis since it will affect only
the exchange of photons with zero energy and momentum, and so would not have 
any effect on the dispersion relation of photons with frequency $\omega>0$.}

Finally, the interaction Lagrangian which is relevant to 
our problem is contained in the third term
in Eq.(\ref{Lmixed}), in particular
\bea
L_{\rm int}^S({\rm \vec{B}})= -\frac{1}{\Lambda_S}\,
\vec{\rm B}\cdot \left(\vec{\nabla}_{x}\wedge \vec{A}(x)\right)\varphi_S(x)
\label{LintSB}
\eea
where the symbol $(a\wedge b)$ indicates the standard vectorial product,
$\vec{\nabla}_{x}\equiv \partial/\partial x_i,~ i=(1,2,3)$
and $\varphi_S(x)$ and $A_{\mu}$ are the scalar and photon fields, 
respectively, with $\vec{A}(x)\equiv A_i(x),~ i=(1,2,3)$.
In momentum space, the corresponding Feynman 
rule associated with the interaction in Eq.(\ref{LintSB}) is given by 
\begin{center}
\epsfbox{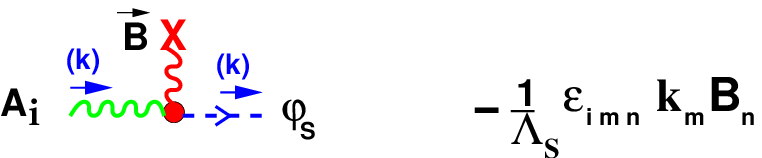}
\end{center}
where $\varepsilon_{i\, m\, n}$ is the antisymmetric tensor in 3 dimensions, 
$k_m$ is the $(m=1,2,3)$ spatial component of the photon 
$k_{\mu}=(\omega,\vec{k})$
entering in the vertex with polarization vector 
$\epsilon_i(k)$.
Notice that at the vertex the energy-momentum is conserved, and that
the external field $\vec{\rm B}$ does not carry any 
four-momentum ($p_{\mu}=0$), since it is assumed to be space-time independent.

\begin{figure}[tpb]
\dofigA{3.1in}{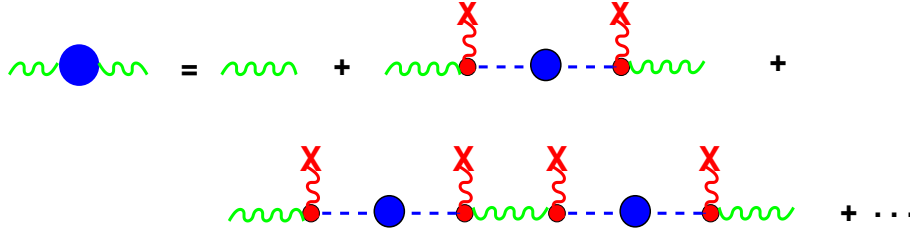}
\caption{\small The effective photon propagator
obtained by summing up
the full series of diagrams with external field insertions 
(curly lines with a cross) carrying zero momentum.
The big bubble in the middle of the dashed lines indicate the 
exact summation of the scalar/pseudoscalar self-energy.}
\label{Self-Energy}
\end{figure}

We are interested in analyzing the modification of the photon 
dispersion relations induced by the interactions in 
Eqs.(\ref{LeffS}),(\ref{LeffP}),
in the presence of a constant and homogeneous external field.
To a first inspection of the leading order effects to the photon 
self-energy, these contributions split in two separate class of diagrams:
\begin{itemize}
\item
the one-loop diagrams with no external field insertions 
induced by the operator $ F_{\mu\nu}(A) F^{\mu\nu}(A) \varphi_S$;
\item
the tree-level diagrams
with two external field insertions induced by the 
mixing operator  $ F_{\mu\nu}^{\rm ext} F^{\mu\nu}(A) \varphi_S$~, 
see Fig.\ref{Self-Energy}.
\end{itemize}
The contribution of the first class of diagrams described above, 
since there are no external field insertions, cannot modify the dispersion
relations. This last effect
can only indeed be achieved by summing up the contribution 
of the second class of self-energy diagrams.

For this reason, in the effective photon propagator, 
obtained by summing up the self-energy diagrams to all orders, 
we do not include the contribution coming from 
the first class of corrections induced by the operator $ F_{\mu\nu}(A) 
F^{\mu\nu}(A) \varphi_S$ to the photon self-energy.
Their effect, as well as the analogous one
of standard QED vacuum polarization diagrams, can be indeed re-absorbed in 
the photon-wave-function renormalization. 
However, there are always higher order contributions 
coming from the mixing corrections induced by
the $ F_{\mu\nu}(A) F^{\mu\nu}(A)\varphi_S$ 
operator and the vertex interaction in Eq(\ref{LintSB}). 
These higher loop effects
should be considered as next to leading order corrections 
to the self-energy contribution induced by
the term in Eq(\ref{LintSB}) and we will neglect them in our analysis.

Now we consider the contributions to the effective photon propagator 
obtained by summing up 
the full series of diagrams as in Fig.\ref{Self-Energy}.
In our calculation we include also the renormalized 
self-energy effects in the scalar propagator, summed up to all orders,
as indicated in Fig.\ref{Self-Energy} with 
a bubble inside the scalar propagator (dashed line).
Notice that the leading contributions to the scalar self-energy
are provided by the one-loop diagrams in which two photons lines are 
circulating in the loop. As shown later on, these class of diagrams 
will be crucial for our analysis since they induce a (finite) 
imaginary part in the scalar self-energy as well as in the effective
photon propagator.

Let us choose a gauge where the free photon propagator $D_{ij}(k)$
in momentum space has only spatial components
\bea
D_{ij}(k)=\frac{i}{k^2+i\varepsilon}\delta_{ij}\, ,
\label{freeprop}
\eea
and zero for all the other ones.
From now on, with the symbol $k^2$ we indicate $k_{\mu} k^ {\mu}$, where
$k_0=\omega$ and $k_i =(\vec{k})_i$ are the corresponding 
energy and 3-momentum, respectively, associated to the photon propagator,
and $\delta_{ij}$ is the unit matrix in 3-dimensional space.
In this gauge, 
the result for the effective photon propagator
$P^{(S,{\rm B})}_{ij}(k,{\rm \vec{B}})$,  induced by the mixing with a 
scalar field 
in the presence of  an external magnetic field $\vec{\rm B}$, is the following
\bea
P_{ij}^{(S,{\rm \vec{B}})}(k,{\rm \vec{B}})&=&
\frac{i}{k^2+i\varepsilon} R_{ij}^{(1)}({\rm \vec{B}})+
\frac{i}{k^2-\Pi^{(1)}(k,{\rm \vec{B}})} 
T_{ij}^{(1)}({\rm \vec{B}})\, ,
\label{propB}
\eea
where 
the tensor 
functions $T_{ij}^{(1)}({\rm \vec{B}})$ and $R_{ij}^{(1)}({\rm \vec{B}})$ 
are symmetric tensors in the $i,j$ indices,
and $T_{ij}^{(1)}({\rm \vec{B}})$ is a 
projector for $\vec{k}$ and ${\rm \vec{B}}$ 
vectors, that is satisfying the conditions
$k_iT_{ij}^{(1)}({\rm \vec{B}})={\rm B}_iT_{ij}^{(1)}({\rm \vec{B}})=0$. 
More explicitly,
\bea
T_{ij}^{(1)}({\rm \vec{B}})&=&\delta_{ij}-\frac{|\vec{\rm B}|^2k_ik_j-
\vec{k}\cdot\vec{\rm B}\left(k_i{\rm B}_j+k_j{\rm B}_i\right)
+|\vec{k}|^2{\rm B}_i{\rm B}_j}{|\vec{k}|^2|\vec{\rm B}|^2-
\left(\vec{k}\cdot\vec{\rm B}\right)^2}\, ,
\nonumber\\
R_{ij}^{(1)}({\rm \vec{B}})&=&
\frac{|\vec{\rm B}|^2k_ik_j-
\vec{k}\cdot\vec{\rm B}\left(k_i{\rm B}_j+k_j{\rm B}_i\right)
+|\vec{k}|^2{\rm B}_i{\rm B}_j}{|\vec{k}|^2|\vec{\rm B}|^2-
\left(\vec{k}\cdot\vec{\rm B}\right)^2}\, ,
\eea
and the expression for scalar part of photon self-energy 
$\Pi^{(1)}(k,{\rm \vec{B}})$
induced by the external magnetic field $\vec{\rm B}$ is given by
\bea
\Pi^{(1)}(k,{\rm \vec{B}})=\frac{1}{|\Lambda_S|^2}\,
\frac{|\vec{k}|^2|\vec{\rm B}|^2\, \sin^2{\theta}}
{k^2-m_S^2-\Sigma_S(k^2)}\, ,
\label{selfE_B}
\eea
where $m_S$ and $\Sigma_S(k^2)$ are the renormalized mass and self-energy
of the exchanged scalar boson, and $\theta$ is the angle between the 
direction of the photon momentum $\vec{k}$ 
and the external magnetic field ${\rm \vec{B}}$.
Notice that the first term on the right hand side of  Eq.(\ref{propB}),
which includes the contribution of longitudinal polarizations, 
depends on the gauge choice in Eq.(\ref{freeprop}), whereas
the second term in  Eq.(\ref{propB}) is gauge independent. 
The latter property comes from the transversality of 
$T_{ij}^{(1)}({\rm \vec{B}})$ tensor, namely 
$k_iT_{ij}^{(1)}({\rm \vec{B}})=0$.

The physical interpretation of 
the $T_{ij}^{(1)}({\rm \vec{B}})$ and $R_{ij}^{(1)}({\rm \vec{B}})$ 
is clear. 
The $R_{ij}^{(1)}({\rm \vec{B}})$ represents the contribution of both  
longitudinal and transverse photon polarization components, 
while $T_{ij}^{(1)}({\rm \vec{B}})$ 
represents the contribution of the transverse polarization 
orthogonal to both $\vec{\rm B}$ and $\vec{k}$.
This picture is largely simplified 
by choosing the momentum $\vec{k}$ direction 
orthogonal to the external field that is 
$\vec{k}\cdot \vec{\rm B}=0$. Then
\bea
T_{ij}^{(1)}({\rm \vec{B}})=\delta_{ij}-\frac{k_ik_j}{|\vec{k}|^2} -
\frac{{\rm B}_i{\rm B}_j}{|\vec{\rm B}|^2}\, ,
~~~~~~~
R_{ij}^{(1)}({\rm \vec{B}})=\frac{k_ik_j}{|\vec{k}|^2}+
\frac{{\rm B}_i{\rm B}_j}{|\vec{\rm B}|^2}\,.
\eea
As expected from the structure of the interaction vertex,
the external magnetic field modifies
the photon dispersion relations only for the transverse polarization component
which is orthogonal to the plane generated by
external magnetic field direction and the photon momentum.
This modification is contained in the denominator of the 
second term of Eq.(\ref{propB}) due to the self-energy
correction $\Pi^{(1)}(k,\vec{{\rm B}})$ in Eq.(\ref{selfE_B}).

Now we consider the case of a scalar field coupled to 
an external electric field ${\rm \vec{E}}$. 
By retaining only the linear terms in the external field,
the relevant interaction Lagrangian is given by
\bea
L_{\rm int}^S({\rm \vec{E}})= -\frac{1}{ \Lambda_S}\,
{\rm \vec{E}}\cdot \left(\frac{\partial}{\partial t}\vec{A}(x)\right)
\varphi_S(x)\, .
\label{LintSE}
\eea
The corresponding Feynman rule is in this case given by
\begin{center}
\epsfbox{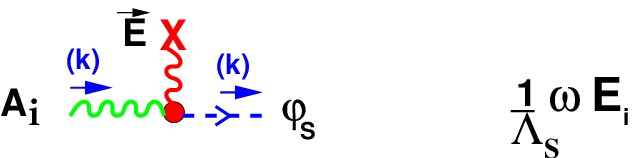}
\end{center}
where ${\rm E}_i$ is the $i$-th component of the external electric field
${\rm \vec{E}}$ and 
$\omega$ 
is the corresponding energy of the photon whose four-momentum 
is entering the vertex.
Notice that the expression of the Lagrangian in Eq.(\ref{LintSE}) 
depends on the adopted choice of temporal gauge $A_0=0$.
However, it is worth stressing that the photon dispersion relations, 
being an observable physical effect, will not depend on the gauge choice.

In analogy with the external 
magnetic field, the effective photon propagator induced by an
external electric field can be straightforwardly derived
by summing up the self-energy diagrams to all orders as above.
By using the same gauge condition 
as in Eq.(\ref{freeprop}) for the free photon propagator,
the result for the effective photon propagator is the following
\bea
P_{ij}^{(S,{\rm \vec{E}})}(k,{\vec E})
&=&\frac{i}{k^2+i\varepsilon} R_{ij}^{(2)}({\rm \vec{E}})+
\frac{i}{k^2-\Pi^{(2)}(k,{\rm \vec{E}})} 
T_{ij}^{(2)}({\rm \vec{E}})\, .
\label{propE}
\eea
where $T_{ij}^{(2)}({\rm \vec{E}})$ and $R_{ij}^{(2)}({\rm \vec{E}})$ 
are given by
\bea
T_{ij}^{(2)}{({\rm \vec{E}})}=\frac{{\rm E}_i{\rm E}_j}{|\vec{\rm E}|^2}\, ,
~~~~~~~
R_{ij}^{(2)}{({\rm \vec{E}})}=\delta_{ij}-
\frac{{\rm E}_i{\rm E}_j}{|\vec{\rm E}|^2}\,.
\eea
The tensor functions 
$T_{ij}^{(2)}{({\rm \vec{E}})}$ and  $R_{ij}^{(2)}{({\rm \vec{E}})}$ 
represent the contributions of the photon polarization
parallel and orthogonal to the external 
electric field ${\rm \vec{E}}$, respectively.
The scalar part of the photon self-energy in the second term of 
Eq.(\ref{propE}) is given by
\bea
\Pi^{(2)}(k,{\rm \vec{E}})=\frac{1}{|\Lambda_S|^2}\,
\frac{\omega^2\, |\vec{\rm E}|^2}{k^2-m_S^2-\Sigma_S(k^2)}\, .
\label{selfE_E0}
\eea
In the case of an angle $\theta\neq \pi/2$ 
formed by the momentum $\vec{k}$ of the
incident wave and the direction of the external electric field $\vec{\rm E}$,
the formula in Eq.(\ref{selfE_E0}) will be modified as 
\bea
\Pi^{(2)}(k,{\rm \vec{E}})=\frac{1}{|\Lambda_S|^2}\,
\frac{\omega^2\, |\vec{\rm E}|^2\sin^2{\theta}}{k^2-m_S^2-\Sigma_S(k^2)}\, .
\label{selfE_E}
\eea
and the corresponding photon polarization will be the one 
parallel to the plane formed by the momentum $\vec{k}$ and the external field
$\vec{\rm E}$.

Contrary to the case of an external magnetic field, 
the external electric field modifies
the photon dispersion relations only for the photon polarization component
which is parallel to the plane formed by $\vec{E}$ and $\vec{k}$, 
while the orthogonal component remains unaffected.
The modification of dispersion relations 
is contained in the denominator of the 
second term of Eq.(\ref{propE}) due to the self-energy
correction $\Pi^{(2)}(k,{\rm \vec{E}})$ in Eq.(\ref{selfE_E}).

Analogous results for a pseudoscalar field minimally 
coupled to two photons as in 
Eq.(\ref{LeffP}) can be obtained in a straightforward way 
from the above results.
In particular, for a pseudoscalar field 
the relevant interaction Lagrangian, linear in the external field, 
these are given by
\bea
L_{\rm int}^{P}({\rm \vec{E}})= \frac{1}{ \Lambda_P}\,
\vec{\rm E}\cdot \left(\vec{\nabla}_{x}\wedge \vec{A}(x)\right)\varphi_P(x)\, ,
\label{LintPB}
\eea
\bea
L_{\rm int}^{P}({\rm \vec{B}})= -\frac{1}{\Lambda_P}\,
{\rm \vec{B}}\cdot \left(\frac{\partial}{\partial t}\vec{A}(x)\right)
\varphi_P(x)\, .
\label{LintPE}
\eea
By comparing the Lagrangians in Eqs.(\ref{LintSB}), (\ref{LintSE}) with
the corresponding ones in  Eqs.(\ref{LintPB}), (\ref{LintPE}) it is easy to 
derive the effective photon propagator by using the previous results for 
the scalar case. In particular, for the effective photon propagator induced
by a pseudoscalar field in the presence of external magnetic or electric field,
one obtains
\bea
P_{ij}^{(P,{\rm {\vec B}})}(k,{\rm {\vec B}})&=&
\frac{i}{k^2+i\varepsilon} R_{ij}^{(2)}({\rm \vec{B}})+
\frac{i}{k^2-\Pi^{(2)}(k,{\rm \vec{B}})} 
T_{ij}^{(2)}({\vec{\rm B}})\, ,
\label{propPB}
\eea
\bea
P_{ij}^{(P,\vec{\rm E})}(k,{\rm {\vec E}})
&=&\frac{i}{k^2+i\varepsilon} R_{ij}^{(1)}({\rm \vec{E}})+
\frac{i}{k^2-\Pi^{(1)}(k,{\rm \vec{E}})} 
T_{ij}^{(1)}({\rm \vec{E}})\, ,
\label{propPE}
\eea
provided that inside the functions
$T_{ij}^{(1)}({\rm \vec{E}})$, $R_{ij}^{(1)}({\rm \vec{E}})$, 
$T_{ij}^{(2)}({\rm \vec{B}})$, $R_{ij}^{(2)}({\rm \vec{B}})$, 
$\Pi^{(2)}(k,{\rm \vec{B}})$, and $\Pi^{(1)}(k,{\rm \vec{E}})$,  appearing in 
Eqs.(\ref{propPB}),(\ref{propPE}), the following substitutions
$\Lambda_S\to \Lambda_P$, 
$m_S\to m_P$, and $\Sigma_S(k^2)\to \Sigma_P(k^2)$
for the corresponding pseudoscalar quantities are implemented.
\section{The photon dispersion}
In order to derive the new dispersion relations for the propagating
photon induced by the external field, we first look at the zeros connected to
the inverse of the real part of photon propagators 
in the last terms of Eqs.(\ref{propB}),
(\ref{propE}) and (\ref{propPB}), (\ref{propPE}).
In particular, this consists in solving
the following {\it mass-gap} equations\footnote{
Even if our {\it mass-gap} equations allow for a massive solution of
the type $k^2=M^2$, this should not be connected to the presence of a phase
transition like in the BCS theory of superconductivity
or in the Nambu Jona-Lasinio model.
A mass term is already present in our Lagrangian and it is given by the
scalar/pseudoscalar mass term. The effective photon propagator can have a
massive pole due to the presence of photon-scalar/pseudoscalar mixing term.
Thus, the mass-gap equations above do not have the same meaning
as in the dynamical mass generation mechanisms mentioned above.}
\begin{itemize}
\item {\bf Scalar }
\bea
{\rm External~~~ \vec{B}}~~~~\to~~~~
 k^2-{\rm Re}\, [\Pi^{(1)}(k,{\rm \vec{B}})]&=&0\, ,
\nonumber \\
{\rm External~~~ \vec{E}}~~~~\to~~~~
k^2-{\rm Re}\, [\Pi^{(2)}(k,{\rm \vec{E}})]&=&0\, ,
\label{selfSB}
\eea
\item {\bf Pseudoscalar}
\bea
{\rm External~~~ \vec{B}}~~~~\to~~~~
 k^2-{\rm Re}\, [\Pi^{(2)}(k,{\rm \vec{B}})]&=&0\, ,
\nonumber \\
{\rm External~~~ \vec{E}}~~~~\to~~~~
k^2-{\rm Re}\, [\Pi^{(1)}(k,{\rm \vec{E}})]&=&0\, .
\label{selfPB}
\eea
\end{itemize}
The fact that there is the presence of an external field,
does not guarantee anymore that the equations above
have a unique solution at $k^2=0$, as 
expected instead in the case of vanishing external sources.

However, we will see that not all the solutions of Eq.(\ref{selfPB})
will correspond to 
physical ones, and there will appear a spurious solution.
In particular, in order to link the solutions to the physical spectrum, 
the following conditions must be fulfilled:
\begin{itemize}
\item
the square of the photon energy $\omega$ must be real and
positive, namely $\omega^2>0$;
\item
the residue at the pole
of the propagator must be positive, in order to ensure that 
the corresponding quantum state has a positive norm. 
\end{itemize}
We will show that the above equations
admit at most 
two {\it acceptable} solutions which satisfy the above criteria.
This can also be seen by switching off the imaginary part of the
scalar/pseudoscalar self-energy. In this case the equations above
become quadratic in $k^2$ and only two solutions are generated, 
corresponding to the classical ones \cite{MPZ}.
One of these solutions corresponds to a light-like 
mode ($\omega=|\vec{k}|/n$) 
with a refraction index $n>1$, while the other one  
can be associated to a {\it  massive} mode ($\omega^2\simeq |\vec{k}|^2+M^2$) 
with a mass $M$ very close to the mass 
of the exchanged scalar or pseudoscalar particle. 
Indeed, the total number of degrees of freedom must be conserved.
In the original Lagrangian there are three
degrees of freedom corresponding to particles which are on-shell, 
the two photon components and the scalar/pseudoscalar field.
The number of physical poles are indeed three, 
the two mentioned above and the pole at $k^2=0$
related to the photon polarization which does not interact with the
external field.

As we will show later on, the inclusion of 
the quantum corrections, which are
absorbed in the scalar/pseudoscalar self-energy, 
cannot be neglected due to the
presence of an imaginary part in the scalar/pseudoscalar self-energy
$\Sigma(k^2)$ 
induced by the interaction in Eqs.(\ref{LeffS}) and (\ref{LeffP}).
Then,
the {\it massive} dispersion mode would be allowed provided that the external 
field is above some particular critical value, which depends on 
the mass and decay width of the exchanged scalar/pseudoscalar particle and
photon energy.

Now we will provide the solutions to the 
{\it mass-gap} equations in Eqs.(\ref{selfSB}), (\ref{selfPB}).
The exact results are given in the appendix A, while
below we will report the approximated expressions
obtained by expanding the exact solutions in powers of a small 
parameter contained in the photon self-energy $\Pi^{(1,2)}(\Delta,k)$.
In order to simplify the notation, it is convenient 
to compactify the set of Eqs.(\ref{selfSB}) and (\ref{selfPB})
as follows
\bea
k^2- {\rm Re}\, \left[\Pi(k,\Delta)\right]=0
\label{gap-eq}
\eea
where $\Pi(k,\Delta)$ generically 
stands for the effective photon self-energy and it is given by
\bea
\Pi(k,\Delta)=
\frac{\Delta}
{k^2-m^2-\Sigma(k^2)}\, .
\label{master}
\eea
Above, $m$ and $\Sigma(k^2)$ indicate 
the renormalized mass and self-energy 
of the exchanged scalar or pseudoscalar particle, respectively.
In Eq.(\ref{master}), the term $\Delta$ absorbs the 
contribution of the external field interaction and it is given by
$\Delta=\Delta^{(1)}({\rm \vec{B}})$ and 
$\Delta=\Delta^{(2)}({\rm \vec{E}})$ for a scalar  field in the 
presence of an external magnetic and electric fields, where
\bea
\Delta^{(1)}({\rm \vec{B}})=\frac{
|\vec{k}|^2|\vec{\rm B}|^2\, \sin^2{\theta}}
{|\Lambda_S|^2}\, ,~~~~
\Delta^{(2)}({\rm \vec{E}})=\frac{
\omega^2\, |\vec{\rm E}|^2\sin^2{\theta}}{
|\Lambda_S|^2 }\,
\label{delta}
\eea
and similarly,  $\Delta=\Delta^{(1)}({\rm \vec{E}})$ 
and $\Delta=\Delta^{(2)}({\rm \vec{B}})$ 
for the pseudoscalar case.

By separating the real and imaginary part in Eq.(\ref{master}), the
Eq.(\ref{gap-eq}) can be written as
\bea
k^2- \frac{\Delta\left(k^2-m^2-{\rm Re} [\Sigma(k^2)]\right)}
{\left(k^2-m^2-{\rm Re}\,[\Sigma(k^2)]\right)^2+\left({\rm Im}\, 
\left[\Sigma(k^2)\right]\right)^2}=0\, .
\label{master2}
\eea
When $k^2=m^2$, the term $m^2-{\rm Re}[\Sigma(m^2)]$ 
corresponds to the renormalized
scalar or pseudoscalar mass, including radiative corrections. However, 
it is convenient to adopt 
the so-called on-shell mass renormalization scheme where
${\rm Re}[\Sigma(k^2)]|_{k^2=m^2}=0$. Then, from now on, 
$m^2$ will stand for the renormalized mass.
On the other hand, ${\rm Im}[\Sigma(k^2)]$ is finite at 1-loop, 
being
connected to the tree level decay in two massless photons. 
In particular,  see Appendix B for more details, 
for scalar (pseudoscalar) interaction we have
\bea
{\rm Im}[\Sigma(k^2)]=-\frac{(k^2)^2}{64\pi\Lambda^2_{S(P)}}\, .
\eea
We stress here 
that the threshold where the 
imaginary part of the scalar/pseudoscalar self-energy 
${\rm Im}[\Sigma(k^2)]$ is different from zero 
starts at $k^2 = 0$, due to the 
fact that a scalar/psedusocalar particle with two photon interaction can always
decay in two massless photons, regardless of the size of its mass.
We will see that the effects of the external field corrections, will 
shift this threshold above $k^2>0$ due to the modification of 
the photon refraction index, see section 4.2 for more details.

In the physical scenarios that we are considering here,
there is a characteristic (small) dimensionless parameter which is given by 
$\Delta/m^4$. Its smallness is 
due to the fact that the $\Delta \sim 1/\Lambda_{S,P}^2$
where the effective scale $\Lambda_{S,P}$
is much larger than  any other characteristic energy scale
present in 
our problem, namely $\Lambda^2_{S,P} \gg \left\{|{\rm \vec{B}}|,
|{\rm \vec{E}}|, \omega^2,m^2\right\}$.
This observation will simplify the analysis
regarding the solutions of Eq.(\ref{master2}).
An approximate solution can indeed be easily found
by using the Taylor expansion in terms of $\Delta/m^4$.
By looking at Eq.(\ref{master2}) one can see that there is 
always a solution around the massless pole $k^2\simeq 0$, which 
can be parametrized as $k^2=\varepsilon_0$, with $\varepsilon_0/m^2$ 
of  order $\Delta/m^4$. This corresponds to the dispersion mode
where the effects of external field 
modifies the refraction index of the vacuum, as for instance in the
case of QED vacuum polarization in the presence of an external magnetic 
field \cite{adler}.
However, as discussed before,
Eq.(\ref{master2}) admits also a {\it massive} solution provided that
$\Delta/({\rm Im}[\Sigma(m^2)])^2 \gg 1$.
As we will show later on, this inequality will imply
the existence of a critical value for the external field.

The Eq.(\ref{master2}), being a cubic equation in $k^2$,
admits three solutions, namely $k^2=M^2_0,~M^2_{\pm}$.
The general expressions are given in the appendix A. 
However, approximated solutions can be easily found 
by using the following antsatz 
\bea
k^2&=&\varepsilon_0\\
k^2&=&m^2+\varepsilon_m\, ,
\eea
By substituting these expressions in Eq.(\ref{master2})
and neglecting higher order terms  ${\cal O}(\varepsilon_{0}^2/m^4)$
and ${\cal O}(\varepsilon_{m}^2/m^4)$, the algebraic equations in $k^2$
can be split into a linear and quadratic ones for 
$\varepsilon_0$ and $\varepsilon_m$, respectively. 
We get the following results 
\bea
M^2_{0}&=&-\frac{\Delta}{m^2}\, ,
\nonumber \\
M^2_{\pm}&=&m^2 + \frac{\Delta}{2m^2}\left(
1 \pm \sqrt{1- \xi}\right)\, ,
\label{gap_sol}
\eea
where 
\bea
\xi ~\equiv~ \frac{4m^6\Gamma^2}{ \Delta^2}\, ,
\label{xidef}
\eea
and $\Gamma$ is the scalar (or pseudoscalar) width given by
${\rm Im}[\Sigma(m^2)]=-m\Gamma$.
As shown in the appendix A, these results can also be
easily obtained from the exact expressions
by retaining the leading order terms in the $\Delta/m^4$ expansion.
In the value of $\Gamma$ we absorb all the effects of the external fields
corrections on the scalar/pseudoscalar decay width.
However, it is reasonable to assume
that these corrections are sub-leading for our problem and 
for practical purposes we neglect them and retain only the
tree-level contributions to the scalar/pseudoscalar width induced by 
the interactions in Eqs.(\ref{LeffS}),(\ref{LeffP}).
In order to have real solutions for $m^2$, the condition $\xi >1$ must be
satisfied. We will return on this point after discussing the dispersion 
relations.

As previously discussed, the lightest mode ($M_0$) 
corresponds to the propagation of a photon in 
a medium with a refraction index $n$ given by
\bea
n^{(S)}({\rm \vec{B}})&\simeq &1+
\frac{|\vec{\rm B}|^2\, \sin^2{\theta}}
{2m^2|\Lambda_S|^2}\\
n^{(S)}({\rm \vec{E}})&\simeq &
1+\frac{
|\vec{\rm E}|^2\, \sin^2{\theta}}{2
m^2|\Lambda_S|^2}\, .
\label{n_refr}
\eea
where $n$ is usually defined as
$n\equiv |\vec{k}|/\omega $.
Above,  $n^{(S)}({\rm \vec{B}})$ 
($n^{(S)}({\rm \vec{E}})$) stands for the
refraction index in magnetic (electric) external fields 
with scalar interactions.
Analogous results are obtained for the pseudoscalar case, with
$n^{(P)}({\rm \vec{B}})=n^{(S)}({\rm \vec{E}})$ and 
$n^{(P)}({\rm \vec{E}})=n^{(S)}({\rm \vec{B}})$\, , where
the substitution of $\Lambda_S\to \Lambda_P$ in 
$n^{(S)}({\rm \vec{B}})$ and $n^{(S)}({\rm \vec{E}})$ is understood.
The next leading order corrections in $\Delta/m^4$ to the dispersion relations
gives the non linear dependence of the refraction index with the
photon energy. From now on in our notation, whenever
the energy ($\omega$) dependence in the refraction index $n(\omega)$ is not 
explicitly shown, it means that it corresponds to $n(\omega=0)$.

Regarding the other {\it massive} solutions $k^ 2=M^2_{\pm}$ of 
Eq.(\ref{master2}), 
only one is physically acceptable and it  will correspond to the pole 
$k^ 2=M^2_{+}$. This can be easily seen
by taking the limit $\Gamma\to 0$  or
analogously $\xi\to 0$, where one should 
recover the classical $(\Gamma=0) $ solution
$k^2=m^2+\frac{\Delta}{m^2}$ \cite{MPZ}.
This {\it massive} pole should correspond to the propagation of a
particle of mass $m=M_{+}$ provided that the square of the mass term
$M^2_{+}$ is a real and positive quantity. 
At this point it is convenient to define the following
dimensionless parameters $x_{E,B}$  by
$x_B=\frac{|\vec{\rm B}|^2\sin^2{\theta}}{2m^2\Lambda_S^2}$
and  $x_E=\frac{|\vec{\rm E}|^2\sin^2{\theta}}{2m^2\Lambda_S^2}$.
In our problem  $x_{E,B}\ll 1 $ and one can simplify the
dispersion relations above by expanding them around  $x_{E,B}=0$. 
In particular, for the scalar case, 
by using the approximate solutions in Eq.(\ref{gap_sol}),
we have for the {\it massive} mode in the presence of external electric 
and magnetic fields 
\begin{itemize}
\item {\bf External~~~ ${\bf \vec{B}}$}
\bea
\omega^2\simeq |\vec{k}|^2\left\{1+x_B\left(1+ \sqrt{1-
\frac{4\Gamma^2m^6\Lambda^4}{|{\rm \vec{B}}|^4\sin^4{\theta}|\vec{k}|^4}}\right)\right\}+m^2\, ,
\label{disp_B}
\eea
\item {\bf External~~~ ${\bf \vec{E}}$}
\bea
\omega^2\simeq \left(|\vec{k}|^2+m^2\right)\left\{
1+x_E\left(1+ \sqrt{1-
\frac{\Gamma^2m^2}{x_E^2\left(|\vec{k}|^2+m^2\right)^2}}\right)\right\}\, ,
\label{disp_E}
\eea
\end{itemize}
where terms of order ${\cal O}(x_{B,E}^2)$ have been neglected.
Analogous results for the pseudoscalar case are obtained by using 
for the external magnetic and electric fields 
the expressions in Eq.(\ref{disp_E}) and  Eq.(\ref{disp_B}), respectively,
and by substituting $\Lambda_S\to \Lambda_P$.

One can easily see that, in order to have a real solution 
for $\omega^2$, see Eq.(\ref{gap_sol}), 
the following condition must be satisfied
\bea
\Delta ~\ge~  2\, m^3\, \Gamma\, .
\label{cond1}
\eea
Notice that in the classical approximation \cite{MPZ}, 
no condition is required for all the parameters in the Lagrangian
in order to have a {\it massive} pole in the spectrum.
Indeed, if one sets $\Gamma\to 0$, the condition (\ref{cond1})
is always satisfied.
However, we emphasize that 
the approximation of neglecting the width $\Gamma$ in the axion physics, as
usually done in all previous studies, is well justified due to the very small
axion mass.
Indeed, for characteristic magnetic fields of 
the order of Tesla, axion masses of the order of meV, and photons 
wave lengths in the optical region, 
the condition (\ref{cond1}) is always satisfied,
provided that the contribution to the  
total width is dominated by the axion decay in two photons.

The physical meaning of Eq.(\ref{cond1})
can be roughly understood as follows.
Let us 
consider the case in which the width $\Gamma$ is very large in comparison
with the mass scale set by the quantity $\Delta$, in such a way 
that the relation
(\ref{cond1}) cannot be satisfied. In this case, the fact that the 
scalar/pseudoscalar particle can decay too fast does not allow 
the {\it massive} photon mode to be formed and hence cannot 
coherently propagate.

Given a particular value for the photon energy $\omega$, 
the inequality above 
leaves to the following critical values for the magnetic and electric 
external fields for a scalar interaction
\bea
|{\rm \vec{B}}| &\ge & {\rm B}_{\rm crit} \equiv 
\frac{m^{\frac{3}{2}}\, \sqrt{2 \Gamma}\, |\Lambda_S|}
{|\vec{k}\sin{\theta}|}\sqrt{1+R(\delta)}
\nonumber \\
|{\rm \vec{E}}| & \ge & {\rm E}_{\rm crit}
\equiv 
\frac{m^{\frac{3}{2}}\, \sqrt{2 \Gamma}\, |\Lambda_S|}
{\omega|\sin{\theta}|}\sqrt{1+R(\delta)}
\label{Bcrit_S}
\eea
and analogously for pseudoscalar one
\bea
|{\rm \vec{E}}| &\ge & {\rm E}_{\rm crit} \equiv 
\frac{m^{\frac{3}{2}}\, \sqrt{2 \Gamma}\, |\Lambda_P|}
{|\vec{k}\sin{\theta}|}\sqrt{1+R(\delta)}
\nonumber \\
|{\rm \vec{B}}| & \ge & {\rm B}_{\rm crit}
\equiv 
\frac{m^{\frac{3}{2}}\, \sqrt{2 \Gamma}\, |\Lambda_P|}
{\omega|\sin{\theta}|}\sqrt{1+R(\delta)}\, .
\label{Bcrit_P}
\eea
where $|\vec{k}|\simeq \sqrt{\omega^2-m^2}
\, +\, {\cal O}(x_{B,E})$. The expression for the function 
$R(x) \sim {\cal O}(x)$, containing the higher-order 
corrections in powers of $\delta=\Gamma/m$,
is reported in the appendix A.
Analogously, for a fixed value of the external field, the conditions
above can be read as the minimum photon energy $\omega$ necessary
in order to generate a {\it massive} mode.
We have explicitly checked that the approximated solutions 
in Eq.(\ref{gap_sol})
are obtained from the exact ones by retaining 
only the leading contribution in the weak external field expansion.
See Appendix A for more details.

A remarkable aspect 
of the results in Eqs.(\ref{Bcrit_S}), (\ref{Bcrit_P})
is that in the case in which the scalar or pseudoscalar
field has only the decay mode in two photons, the critical value for the
external field does not depend on the coupling $\Lambda$ 
of the effective interaction. This can be easily checked 
by noticing that the corresponding width $\Gamma$ in that case would be
proportional to $\Gamma\propto m^3/\Lambda^2$, while
$\Delta\propto 1/\Lambda^2$.

In order to identify the {\it massive} poles in the effective propagator 
as physical quantum states,
one has to check that 
the corresponding residue at the pole of the propagator is positive.
Indeed, the residue at the pole ($Z$) is connected by unitarity 
to the norm of the quantum state excited from the vacuum.
In particular, 
for a generic solution $k^2=M^2$ of the mass gap equation, one gets
\bea
Z\equiv \lim_{k^2\to {\rm M}^2}\left(k^2-{\rm M}^2\right)\left(k^2-
{\rm Re}\, \Pi(k,\Delta)\right)^{-1}\, .
\eea
or equivalently
\bea
Z=\left(1-\frac{\partial}{\partial k^2} {\rm Re} \, \Pi(k,\Delta)
\Big|_{k^2={\rm M}^2}
\right)^{-1}\, .
\label{Z}
\eea
where $Z^{1/2}$ is the wave function renormalization constant.

By using the expression in Eq.(\ref{Z}),
we find the following result for the residue at the poles $k^2=M_0^2$ 
and $k^2=M_{\pm}^2$, respectively $Z_{0}$ and  $Z_{\pm}$
\bea
Z_{0}&=&
1-\frac{\Delta}{m^4}\, , 
\\
Z_{\pm}
&=&\pm\, 
\frac{\Delta}{2m^4}\frac{\left(1 \pm\sqrt{1-\xi}\right)}{\sqrt{1-\xi}}\, ,
\label{res}
\eea
where we have neglected terms of the order
${\cal O}(\frac{\Delta^2}{m^8})$.
As we can see from these results, two physical solutions are allowed,
corresponding to the positive norm states of $Z_0>0$ and $Z_+>0$, 
connected respectively to the poles $k^2=M^2_{0}$
and $k^2=M^2_{+}$, while $k^2=M^2_{-}$ is an unphysical one,
being associated to a ghost ($Z_{-} <0$).

Notice that when the external field is far above the
critical value, $\xi\ll 1$,  but still in the weak coupling regime
i.e. $\Delta/m^4\ll 1$, then
the term $Z_{+}\simeq {\cal O}(\Delta/m^4) \ll 1$.
This means that in this case 
the probability to induce the {\it massive} mode from the
vacuum, is very suppressed. 
However, when we approach the critical value
$\xi\to 1$ the constant $Z_{+}$ increases and tends to infinity at
$\xi=1$. 

This behavior can be easily understood if we look 
at the inverse of the propagator $\Pi^{-1}(k^2)$ 
near the pole $k^2\simeq M^2$, in particular
\bea
\Pi^{-1}(k^2)|_{k^2\simeq M^2} = (k^2-M^2) Z^{-1} + 
(k^2-M^2)^2 \, {\rm Re}\Sigma^{''}(M^2)|_{k^2=M^2} 
+ \dots \, +i\,{\rm Im}\Sigma(M^2)
\eea
where the dots stand for higher order terms in $(k^2-M^2)/m^2$ expansion, 
and $\Sigma^{''}(M^2)$ indicates the second derivative 
of $\Sigma(k^2)$ with respect to $k^2$ and evaluated at $k^2=M^2$.
Notice that $\xi=1$ corresponds to two degenerate {\it massive} 
solutions, 
and the propagator generates a double pole in $k^2=M^2_{+}=M^2_{-}$. 
This is not a matter of the approximation adopted, since the
same conclusions are obtained in the exact case
(see Appendix A for more details).
Clearly, 
since $Z^{-1}$ is related to the coefficient of the single pole, 
$Z^{-1}$ vanishes at  $\xi\to 1$ or analogously $Z\to \infty$.
Moreover, since the $Z$ factor is related by unitarity to the probability of 
creating the corresponding quantum state from the vacuum, when 
$Z>1$ the unitarity is spoiled. Therefore, the requirement of 
unitarity sets a region of validity of our calculations
when the external field is close to its critical value.
By imposing $Z_{+}\le 1$ we obtain
\bea
0\le \xi\le 1-\frac{\Gamma^2}{m^2}+{\cal O}(\frac{\Gamma^3}{m^3})\,.
\eea
In conclusion, the probability that the {\it massive} mode is 
excited from the vacuum increases as the external field
approaches (from above) the critical value, but clearly vanishes
just below the critical value. 

\section{The photon absorption coefficient}
Due to the presence of an external constant electric or magnetic field,
the interactions in Eqs.(\ref{LeffS}), (\ref{LeffP})
can induce a non-vanishing probability for the photon splitting amplitude
$\gamma\to \gamma \gamma$. This phenomenon is well known in 
the framework of  QED \cite{adler},  as well as the analogous one of
electron-pair  creation $\gamma\to e^ +e^ -$ in constant
and homogeneous magnetic field \cite{pair_creat,erber}.
Nevertheless, there are 
no studies so far concerning the analogous effect of 
photon conversion process $\gamma\to \gamma\gamma$
induced by the scalar and pseudoscalar 
interactions.\footnote{However, there is an analogous
study analyzing the effects of external field on 
photon and axion $(a)$ decays $\gamma\to a \gamma$ and  
$a\to \gamma \gamma$ in the framework of a very 
light axion \cite{axion3}. We stress that the analysis and results 
contained in \cite{axion3} are quite different from the ones
presented in our work and there is not any significant overlap.}
Measurement of a photon splitting in constant magnetic field is
a challenge for experiment, although 
high-energy photon splitting in atomic fields has been recently observed 
\cite{splitting_exp}.

For this kind of problem it is appropriate to express
the conversion probability in terms of a photon attenuation
coefficient, usually called {\it absorption} coefficient ($d_{\gamma}$)
\cite{erber}.
In particular, the number $N_{\gamma\gamma}$  of photon splitting
events  created
by a photon crossing an EM background field for a path length $L$
are given by
\bea
N_{\gamma\gamma}=N_{\gamma}\left(1-\exp\left[-d_{\gamma}\, L\right]\right)
\simeq \, N_{\gamma} d_{\gamma}\, L
\label{d_gamma}
\eea
where $N_{\gamma}$ is the total number of photons entering the 
background EM field. Notice that the last relation 
is a good approximation only when $d_{\gamma}L\ll 1$.

Due to the photon interactions in Eqs.(\ref{LeffS}), (\ref{LeffP}),  
a new kind of contribution to the photon absorption
coefficient $d_{\gamma}$ is expected with respect to the 
corresponding vacuum polarization effects in QED \cite{adler,pair_creat,erber}.
Now, before entering into the analysis of the new physics contributions,
let us start by  recalling the known results on the 
photon propagating in a constant magnetic field \cite{adler}. 

In QED, the matrix element for the $\gamma\to \gamma \gamma$
process can be calculated in perturbation theory by using the 
Euler-Heisenberg effective Lagrangian \cite{EH} arising from
integrating out electrons at 1-loop.
After expanding the  electromagnetic field around the constant background
field, new dispersions
relations are obtained for the photon propagating in the 
external field. A refraction index $(n_{(\lambda)} > 1)$ 
associated to  the photon polarization $\varepsilon_{\mu}(\lambda)$, 
where $\lambda$ parametrizes the two polarizations parallel and 
perpendicular to the magnetic field direction,
arises.
Even 
by neglecting the dispersion effects, the photon splitting process
$\gamma\to \gamma \gamma$ is kinematically allowed, 
provided that all particle momenta in the reaction are proportional
to a unique momentum $k_{\mu}$, satisfying $k^2=0$.
However, as shown in Ref.\cite{adler}, at least 
three external field insertions would be necessary in order to have 
a non-vanishing matrix element.
This is a simple consequence of the Lorentz and gauge invariant
structure of the Euler-Heisenberg Lagrangian and due to the fact 
that there is only one light-like four-momentum in the process.
The magnitude of the resulting photon splitting 
absorption coefficient would then given by  \cite{adler}:
\bea
d_{\gamma}^{\rm QED}\simeq \, 0.1\left(\frac{{\rm B}}{
{\rm B}_{\rm cr}^{\rm QED}}\right)^6\left(
\frac{\omega}{m_e}\right)^5 \, {\rm cm}^{-1}\, ,
\eea
where  ${\rm B}_{\rm cr}^{\rm QED} =\frac{m_e^2}{e}$, with $e$ the unity of 
electric charge, is the critical magnetic field for the
photon-pair creation \cite{pair_creat,erber} and $m_e$ is the
electron mass. This formula holds only for $\theta=\pi/2$ and 
in the weak-field limit $e{\rm B}/m_e^2\ll 1$.
Then, one can see that even for strong 
laboratory magnetic fields (of the order of Tesla) the
absorption coefficient  $d_{\gamma}$ would be very small, due to the
fact that $B_{\rm cr}^{\rm QED} \sim 10^{10}$ Tesla.
Only in
astrophysical context this effect can become relevant. 
In particular, with pulsar magnetic fields of the order of 
$B_{\rm cr}^{\rm QED}$ and $\omega\simeq m_e$, there are many photon splitting 
absorption lengths in a characteristic pulsar distance of $10^6$cm.
By taking into account the dispersive effects, the energy-momentum 
conservation is modified. The momenta of  the two final photons
will not be parallel anymore and differ from each other for small angles.
Due to the dispersive effects, the matrix element
of the photon splitting can be induced by
one external field insertion. However, in this case
the corresponding $d_{\gamma}$ would turn out very suppressed 
by kinematical factors being proportional to the small opening angle
of the final photons.
\cite{adler}.

Returning to our case, a practical way to calculate the
absorption  coefficient from the photon self-energy
in Eq.(\ref{master}) is 
by making use of the optical theorem which connects $d_{\gamma}$
to the imaginary part of the self-energy evaluated at the 
corresponding poles of the propagator.
As can be seen from Eq.(\ref{selfE_B}), the imaginary part
of photon self-energy ${\rm Im}\Pi(k,\Delta)$ is then 
proportional to the imaginary part of the scalar/pseudoscalar self-energy,
namely ${\rm Im}\,  \Pi(k,\Delta) \propto {\rm Im} \Sigma(k^2)$.

Now, if we do not take into account the effects of vacuum polarization
in background EM field, 
and consider the photon purely massless, the kinematics of the
reaction  $\gamma\to \gamma \gamma$ 
would not forbid this process provided that all the momenta are proportional
to a unique light-like momentum $k^2=0$.
However, as in the
analogous case of QED, due to the fact that there is only one independent 
light-like four momentum and due to the antisymmetric property of the 
electromagnetic field strength $F_{\mu\nu}$, the minimum 
number of external field insertions is equal to three 
in order to have a non vanishing effect in the matrix element.
This is a general result,
as proved by Adler \cite{adler}, and does not depend on the structure of the 
interaction, but on the gauge and Lorentz 
invariant property of the effective Lagrangian
for the photon obtained after integrating out the other 
degrees of freedom.
In the case of weak magnetic fields, this 
would lead into a very strong suppression.
As we will show later on, 
if dispersion effects are taken into account, ($k^2=M_0^2$), 
the process could proceed by means of one external field insertion only,
but suppressed by the small angle induced by the dispersions effects.
In the next two sub-sections we will analyze the 
photon splitting phenomenon 
induced by the {\it massive} and massless dispersion modes 
respectively.

\subsection{Massive dispersion mode}
As shown in the previous section, the 
{\it massive} mode of the photon, with mass of the order of 
the scalar/pseudoscalar one ($m$), 
is allowed by dispersion relations when the
critical conditions in Eqs.(\ref{Bcrit_S}),
(\ref{Bcrit_P}) are satisfied. Then, the {\it massive} mode 
could easily decay in two lighter photons, or in 
any other kinematically allowed final states $f$ 
coupled to the intermediate scalar/pseudoscalar particles.
In this case, the corresponding 
{\rm width} will be proportional to
the imaginary part of self-energy evaluated on the
{\it massive} pole of the photon propagator ($k^2=M^2_{+}$).
In conclusion, the photon 
could get a non-vanishing decay 
{\it width} ($\Gamma_{\gamma}$), provided that
the critical conditions in Eqs.(\ref{Bcrit_S}),
(\ref{Bcrit_P}) are satisfied and $\omega \gsim  m_{\gamma}$, where
$m_{\gamma}\equiv \sqrt{M^2_{+}}$.

One can formally define a {\it width} ($\Gamma_{\gamma}$)
associated to the {\it massive} mode by making use of the optical theorem. 
In particular, 
\bea
m_{\gamma}\, \Gamma_{\gamma}=-Z_{+}\, 
{\rm Im}\left[\Pi(k,\Delta)\right]\Big|_{k^2=m^2_{\gamma}}
\theta(\omega-m_{\gamma})\theta({\rm B}-{\rm B}_{\rm crit})
\, ,
\label{imP}
\eea
where $\theta(x)$ is the standard $\theta$-function defined as
$\theta(x) =1$ for $x\ge 0$ and $\theta(x) =0$ for  $x< 0$, 
and $m_{\gamma}
=\sqrt{m^2+
\frac{\Delta}{2m^2}(1+\sqrt{1-\xi})}$. The constant $Z_{+}$ appearing 
in Eq.(\ref{imP})
is the corresponding residue at the pole, given in Eq.(\ref{Z}).
As shown later, the same result for $\Gamma_{\gamma}$ 
in Eq.(\ref{imP})
can be re-obtained by starting from the matrix element of the transition 
$\gamma^{*}\to \gamma \gamma$,
where $\gamma^{*}$ represents the massive mode of the photon.
For the imaginary part of photon self-energy evaluated at the {\it massive} 
pole $k^2=m_{\gamma}^2$ we have
\bea
{\rm Im}\left[\Pi(k,\Delta)\right]|_{k^2=m_{\gamma}^2}
=-m^2\frac{\left(1-\sqrt{1-\xi}\right)}{\sqrt{\xi}}\, .
\label{Imsigma}
\eea
Then, the  $\Gamma_{\gamma}$ is given by
\bea
\Gamma_{\gamma}=\frac{\Gamma}{\sqrt{1-\xi}}\left(1-
\frac{\Gamma}{2m}\frac{\left(1+\sqrt{1-\xi}\right)}{\sqrt{\xi}}\right)\, ,
\eea
where the last term in parenthesis comes from the first order expansion 
in $\Delta/m^4$ of the $m_{\gamma}$ when expressed in terms of the 
axion mass $m$.
The absorption coefficient $d_{\gamma}$ entering in Eq.(\ref{d_gamma}), 
is related to the $\Gamma_{\gamma}$ in Eq.(\ref{imP}) by
\bea
d_{\gamma}=\frac{m_{\gamma}}{\omega}\, \Gamma_{\gamma}\, .
\label{dgamma_f}
\eea
Finally, by inserting the results of Eqs. (\ref{res}), 
(\ref{imP}), and (\ref{Imsigma}) into Eq. (\ref{dgamma_f}), we obtain
\bea
d_{\gamma}&=&\frac{\Gamma}{\sqrt{1-\xi}}\frac{m}{\omega}
\, 
\label{dgamma}
\eea
where we neglected terms of the order ${\cal O}(\Gamma^2/m^2)$. 
We stress that 
to be more precise the scalar/pseudoscalar width $\Gamma$ appearing 
in all the calculations above 
should be the one evaluated on the massive pole $m_{\gamma}$, that is 
$\Gamma(m_{\gamma})$, which differs from the width $\Gamma$ 
evaluated on the scalar/pseudoscalar mass-shell mass $m$ 
by small terms of order ${\cal O}(\Gamma/m)$.
The same argument applies for the mass $m$ appearing in the expression 
$m\Gamma$ which should be $m_{\gamma}$ rather than $m$.

As explained at the end of the previous section, the
restriction on the upper limit 
$\xi^{\rm max}= 1-\frac{\Gamma^2}{m^2}$ comes from
the requirement of unitarity $Z_{+}\le 1$. Indeed, the validity of the 
results in Eq.(\ref{dgamma}) is based on the optical theorem, 
which holds only under the hypothesis of unitarity. Notice that, for the
imaginary part of photon self-energy evaluated on the 
other pole $k^2=M_{-}^2$, the corresponding absorption coefficient 
would have been negative due to the fact that $Z_{-}<0$, 
pointing out the presence of an unphysical solution.

These results can be easily re-obtained by starting from the standard
formula for the absorption coefficient \cite{adler} 
of photon splitting $\gamma(\omega)\to \gamma(\omega_1)\gamma(\omega_2)$
\bea
{\rm d} d_{\gamma}=\frac{1}{2\omega}\frac{1}{2}
\frac{d^3 k_1}{(2\pi)^3 2\omega_1}
\frac{d^3 k_2}{(2\pi)^3 2\omega_2} (2\pi)^4\delta^4(k-k_1-k_2)\, 
|M(\gamma\to \gamma \gamma)|^2
\label{ddg}
\eea
where
$|M(\gamma\to \gamma \gamma)|^2$ is the square modulus 
summed over final state polarization of the corresponding amplitude. 
The extra factor $1/2$ in front, takes into account for the identical final 
states of two photons.
The amplitude at the leading order can be easily obtained by evaluating 
the following Feynman diagram
\bea
\epsfbox{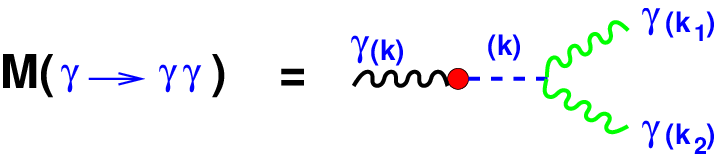}
\label{Mgg}
\eea
where the dark and light curly lines represent the massive and massless
photon propagation modes of the photon, and
the bubble stands for the photon-scalar/pseudoscalar vertex
induced by the external field,
and the dashed line represents the scalar/pseudoscalar propagator.
By taking into account the renormalization of the wave function 
$Z_{+}$ in the massive mode, as given by Eq.(\ref{res}), one obtains
\bea
|M(\gamma\to \gamma \gamma)|^2=\frac{Z_{+}\Delta}{(k^2-m^2)^2
+\Gamma^2m^2}\sum_{\rm pol} |V_{\gamma\gamma}(k_1,k_2)|^2
\label{M2}
\eea
where $V_{\gamma\gamma}(k_1,k_2)$ stands for the vertex of 
scalar/pseudoscalar in two photons and the
sum is performed over the final photon polarizations.
In particular, for the scalar (pseudoscalar) interactions,
$V_{\gamma\gamma}(k_1,k_2)=V^{S(P)}_{\gamma\gamma}(k_1,k_2)$, with
\bea
V^{S}_{\gamma\gamma}(k_1,k_2)&=&\frac{1}{2\Lambda_S}
\left[(k_1\cdot k_2)g^{\mu\nu}-k_1^{\mu} k_2^{\nu}\right]
\varepsilon_{1\nu}(k_1)\varepsilon_{2\mu}(k_2)\, ,
\nonumber
\\
V^{P}_{\gamma\gamma}(k_1,k_2)&=&\frac{1}{\Lambda_P}
\left(k_1^{\mu}\, \varepsilon_1^{\nu}\, 
k_2^{\alpha}\, \varepsilon_2^{\beta}\right)\epsilon_{\mu\nu\alpha\beta}\, ,
\label{Vertex}
\eea
where $\varepsilon_1^{\mu}(k_1)$ and $\varepsilon_2^{\mu}(k_3)$
are the polarization vectors of the two final photons, and $k_{1,2}$ the 
corresponding four momenta.
Then, by evaluating $k^2$ on the massive pole
$k^2=m_{\gamma}^2$, Eq.(\ref{ddg}) can be re-written as 
\bea
{\rm d} d_{\gamma}=
\frac{Z_{+}\Delta}{(m_{\gamma}^2-m^2)^2
+\Gamma^2m^2}
\left\{d\Phi (2\pi)^4\delta^4(k-k_1-k_2)\, 
\sum_{\rm pol} |V_{\gamma\gamma}(k_1,k_2)|^2
\right\}\, ,
\label{dg2}
\eea
where $d \Phi \equiv
\frac{1}{2\omega}\frac{1}{2}
\frac{d^3 k_1}{(2\pi)^3 2\omega_1}
\frac{d^3 k_2}{(2\pi)^3 2\omega_2}$.
Notice that the integral over the phase space in the
last term in parenthesis gives just the scalar/pseudoscalar width 
$\frac{m_{\gamma}}{\omega}\Gamma(m_{\gamma})\simeq \frac{m}{\omega}\Gamma$.
Finally, by integrating  Eq.(\ref{dg2}) and using the identity
\bea
\frac{Z_{+}\Delta}{(m_{\gamma}^2-m^2)^2+\Gamma^2m^2}=\frac{1}{\sqrt{1-\xi}}\, ,
\eea
one can easily recover Eq.(\ref{dgamma}).
It is worth noticing that 
the denominator of the right-hand-side (r.h.s.) of Eq.(\ref{dg2}), which is
connected to the 
scalar(pseudoscalar) propagator, is going in resonance since $m_\gamma$ is
very close to $m$. This effect partially removes the suppression
given by the $Z_+$ in the numerator, 
that is the probability to induce the massive mode.

At this point it is fair to say that the results obtained for 
the absorption coefficient are based on the assumption that the
EM background field is constant and homogeneous in all space.
This is an approximation since for any 
practical experiment the external magnetic or electric field 
has finite extension.
However, it is reasonable to believe that the effects of 
the boundary conditions can be neglected (as in our case) when the 
extension length $L$ of the external EM background field where 
the photon is traveling satisfies the condition $L\gg \lambda_{{\gamma}}$,
where $\lambda_{{\gamma}}\sim 1/m_{\gamma}$ 
is the de Broglie wave length associated to the {\it massive} mode
$k^2=m_{\gamma}^2$.

Now we analyze two particular limiting cases, the region
of external fields very close to the critical value
$\xi\simeq 1$, and the large external fields $\xi\ll 1$.
It is easy to see that 
the maximum value of the width, compatible with unitarity,
is obtained, as expected, near the resonant region $\xi\to 1$
where the $Z_{+}\to \infty$. 
By restricting ourself to the region of maximum value of $Z_{+}$ 
allowed by unitarity, that is  $Z_{+}\simeq 1$ and corresponding to 
$\xi\simeq 1-\Gamma^2/m^2$,
the maximum value of the absorption coefficient is
\bea
d_{\gamma}^{\rm max}=\frac{m^2}{\omega}
+{\cal O}(\frac{\Gamma}{m})\, .
\label{imPmax}
\eea
In the opposite limit, 
of very large external fields $\xi\to 0$, the absorption 
coefficient gets its minimum value given by
\bea
d^{\rm min}_{\gamma}\, \simeq\,  \frac{m\Gamma}{\omega}\,
\label{imPlimit}
\eea
which is independent of the external field.

\begin{figure}[tpb]
\dofig{3.1in}{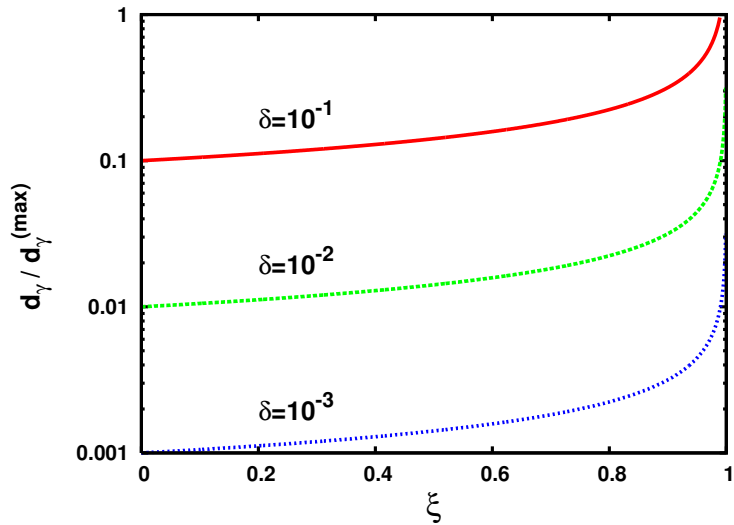}
\caption{\small The ratio 
$d_{\gamma}/d_{\gamma}^{\rm max}$, versus $\xi$, for three values
of $\delta=\Gamma/m$, with $\delta=10^{-1},10^{-2},10^{-3}$.
 }
\label{R}
\end{figure}

In Fig. \ref{R} we plot the function 
$d_{\gamma}$ normalized to its maximum value 
$d_{\gamma}^{\rm max}$, versus $\xi$, for some representative 
values of ratios $\Gamma/m$. 
As can be seen from these results, the variation of the photon absorption 
coefficient as a function of $\xi$ (or analogously as a function of the 
external field if all the other parameters are taken constants)
is very steep only near the region close 
to the maximum value $\xi^{\rm max}=1-\Gamma^2/m^2$, when $\Gamma/m\ll 1$.
Therefore, for very small values of $\Gamma/m$ 
the average of $d_{\gamma}$
over the external field could provide a good estimation of the
size of the effect, in particular
\bea
\bar{d}_{\gamma}\,\simeq\,\int_{0}^{1-\frac{\Gamma^2}{m^2}} d\xi~ 
d_{\gamma}=\,\frac{2\Gamma \, m}{\omega}\left(1-\frac{\Gamma}{m}\right)
+{\cal O}(\frac{\Gamma^2}{m^2})\, .
\label{dbar}
\eea
However, notice that the average over the full available range of $\xi$,
($0\le \xi \le 1$), is finite and given by
$\bar{d}_{\gamma}=
\,\frac{2\Gamma \, m}{\omega}(1+{\cal O}(\frac{\Gamma^2}{m^2}))$. 
The difference between this result and the one obtained 
in Eq.(\ref{dbar}), with the restricted integration region
is a quantity of order ${\cal O}(\Gamma/m)$.
This difference should roughly give an idea
of the theoretical uncertainty on the
absorption coefficient when the average over the external field is 
considered. 

\subsection{Massless dispersion mode}
Here we consider the possibility that the photon splitting in two photons is
induced by the massless dispersion mode $k^2=M_0^2$, see Eqs.(\ref{gap_sol})
and (\ref{M0_exact}). 
The transition amplitude is given by the Feynman diagram as
in Eq.(\ref{Mgg}), where now 
the initial photon dispersion corresponds to the massless mode.
The evaluation of the absorption coefficient can be easily performed 
by using the standard formula in Eq.(\ref{ddg}), 
as in the case of QED \cite{adler}, provided that dispersion relations
for the final photons are taken into account.
At this point we should stress that if dispersion effects are neglected in 
the initial and/or final photons propagation modes (that is $n=1$), 
then the corresponding 
absorption coefficient vanishes due to the Lorentz covariant
structure of the vertex 
$V_{\gamma\gamma}(k_1,k_2)$ in Eq.(\ref{Vertex}).
Indeed, although the process of photon splitting with 
no-dispersion is not forbidden 
by kinematics provided that all momenta in the reaction
are proportional to a unique momentum satisfying 
$k^2=0$, the square amplitude 
$\sum_{\rm pol} |V_{\gamma\gamma}(k_1,k_2)|^2\propto (k_1\cdot k_2)^2$ 
vanishes when $k_1^2=k_2^2=k^2=0$.

The standard formula for the absorption coefficient
remains the same as in Eq.(\ref{ddg}) if the
dispersion effects are implemented in the delta function as \cite{adler}
\bea
\delta^{4}(k-k_1-k_2)\to \delta(\omega -\omega_1-\omega_2)
\delta^{3}(
n(\omega)\, \omega \hat{k}-n_1(\omega_1)\, 
\omega_1 \hat{k}_1-n_2(\omega_2)\, \omega \hat{k}_2)\, ,
\label{ph_space}
\eea
where $\omega, \omega_{1,2}$ and $k,k_{1,2}$ are the energy and modulus of the 
momenta of the corresponding waves respectively, $\hat{k}=\vec{k}/|\vec{k}|$ 
is the unity vector associated to the momentum $\vec{k}$, 
and generically $n(\omega)$ is
the refraction index where the explicit dependence on the wave energy 
$\omega$ is shown.
According to Ref. \cite{adler}, 
the condition set by kinematics in order the reaction to proceed is 
\bea
\Sigma&\equiv& \omega_1 n_1(\omega_1) + \omega_2 n_2(\omega_2)
-(\omega_1+\omega_2) n(\omega_1+\omega_2)\,  >\, 0\, .
\label{kin_cond}
\eea
This condition can be easily understood by looking at the
opening angle $\theta_{12}$
between the two 3-momenta $\vec{k_1}$ and $\vec{k_2}$ of the final 
photons, where
\cite{adler}
\bea
\theta_{12}=\frac{\omega}{\left(\omega_1\omega_2\right)^{1/2}}
\left(\frac{2\Sigma}{\omega}\right)^{1/2}\, .
\label{theta12}
\eea
In order to have real values for $\theta_{12}$, the
condition $\Sigma\ge 0$ must be satisfied.

Let us start by analyzing the case of a photon crossing a constant and
homogeneous magnetic field in the presence of a scalar interaction.
In order to simplify the analysis we consider the 
case of a linearly polarized wave with momentum $\vec{k}$
perpendicular to the magnetic field direction.
As a first approximation, we neglect the effects of the external field 
on the QED vacuum polarization.
As we will show later on, this approximation can be well 
justified for typical mass and coupling $\Lambda$ 
in the range of  $m\sim  10^{-3}$eV and 
$\Lambda\sim 10^6$GeV, as suggested for instance by the results of PVLAS
experiment \cite{PVLAS},
or in general for $\sqrt{m\Lambda} \lsim 5\times m_e \sim 2.5 {\rm MeV}$, 
where $m_e$ is the electron mass.

From now on in this section we will use the following notation for the
polarizations of the photons:
namely $\gamma_{\parallel}$ ($\gamma_{\perp}$), corresponding to the 
polarization vector $\vec{\epsilon}_{\parallel(\perp)}(k)$ 
parallel (perpendicular) to the plane formed by the external field 
direction $\vec{{\rm B}}$ and the direction of wave propagation
$\hat{k}=\vec{k}/|\vec{k}|$, where $\vec{k}\cdot\vec{\varepsilon}(k)=0$.
Moreover, $n_{\parallel}(\omega)$ and  $n_{\perp}(\omega)$ stand for the 
corresponding refraction indices.
One can first notice that, in the case of scalar interaction, 
only the $\gamma_{\perp}$ will acquire
a refraction index $n >1$, if the direction of $\vec{k}$ is 
perpendicular to $\vec{{\rm B}}$.

Let us consider the dispersion relations for the polarization state
$\gamma_{\perp}$ with refraction index $n_{\perp}(\omega)$
in the case of weak magnetic fields, or small frequencies, where
$\Delta/m^4\ll 1$. At this point 
it is convenient to introduce an energy
scale defined as $\mu\equiv \sqrt{m\Lambda}$. As will be clear later on, 
this scale plays the role of an effective energy scale, as
the analogous of the electron mass $m_e$ in the QED 
photon splitting phenomenon.
From Eq.(\ref{delta}), the term $\Delta$ (for $\theta=\pi/2$) is given by
$ \Delta=|\vec{k}|^2 \left(\frac{{\rm B}}{{\rm \tilde{B}}_{\rm cr}}\right)^2$
where ${\rm \tilde{B}}_{\rm cr}\equiv \mu^2$. \footnote{
Note that the critical value of the external
magnetic field ${\rm \tilde{B}}_{\rm cr}$ associated to the massless
mode is different from the critical magnetic field 
connected to the generation of massive mode, namely ${\rm B}_{\rm crit}$,
in Eqs.(\ref{Bcrit_S}),(\ref{Bcrit_P}).}
The solution of $k^2=M_0^2$ can be expanded in powers of photon energies
$\omega$, provided that $\omega\ll \mu$. By using the 
results reported in Appendix A, the
dispersion relations at the next to leading order expansion in $\Delta/m^4$
give in this case
\bea
\omega^2=|\vec{k}|^2\left(1-\left(\frac{{\rm B}}{{\rm \tilde{B}}_{\rm cr}}\right)^2
+\frac{|\vec{k}|^2}{m^2}\left(\frac{{\rm B}}{{\rm \tilde{B}}_{\rm cr}}\right)^4
\right)
+{\cal O}(\left(\frac{\omega {\rm B}}{m {\rm \tilde{B}}_{\rm cr}}\right)^6)\, .
\eea
The index $n_{\perp}(\omega)$ 
is then easy to calculate and it is given by
\bea
n_{\perp}(\omega)=
\left(1-\left(\frac{{\rm B}}{{\rm \tilde{B}}_{\rm cr}}\right)^2
+\frac{\omega^2}{m^2} \left(\frac{{\rm B}}{{\rm \tilde{B}}_{\rm cr}}\right)^4
\right)^{-1/2}
\label{n_perp}
\eea
where we neglected higher order terms in the 
$\frac{{\rm B}}{{\rm \tilde{B}}_{\rm cr}}$ expansion. 
Three possible kinds of transitions for polarized states are allowed
\bea
\gamma_{\perp}&\to& \gamma_{\perp} + \gamma_{\perp}
\\
\gamma_{\perp}&\to& \gamma_{\parallel} + \gamma_{\parallel}
\\
\gamma_{\perp}&\to& \gamma_{\perp} + \gamma_{\parallel}\, .
\eea
In order to see if the reaction can proceed, one has to first check 
if the kinematical condition in Eq.(\ref{kin_cond}) is satisfied.
For this purpose it is
useful to adopt the following approximation. Since the angles between 
the final  and initial photon directions of propagation are very small, 
all the final momenta are almost aligned along the initial momentum. 
This implies that the modification of the refraction 
index $n_{\parallel}$ corresponding to 
the parallel polarization wave $\gamma_{\parallel}$ of the final photon 
is much smaller than the corresponding one in the other polarization
$\gamma_{\perp}$, being suppressed by the
small opening angle. In particular this means that
$n_{\parallel}(0)-n_{\perp}(0) < 0$.
Therefore, for the refraction indices of final photons
we can approximate $n_{\parallel}=1$, 
while keeping $n_{\perp}(\omega)$ 
as the one in Eq.(\ref{n_perp}).
By evaluating the expression $\Sigma$ using the Taylor expansion  
up to the second order in the photon frequency \cite{adler}, 
we get 
\bea
\Sigma[\gamma_{\perp}\to (\gamma_{\perp})_1 
(\gamma_{\perp})_2]
&=&-\frac{3}{2}n^{''}_{\perp}(0)\left(\omega_1^2\omega_2
+\omega_1\omega_2^2\right) > 0
\nonumber\\
\Sigma[\gamma_{\perp}\to (\gamma_{\parallel})_1 
(\gamma_{\parallel})_2]
&=&(\omega_1+\omega_2)\left(
n_{\parallel}(0)-n_{\perp}(0)\right) < 0\, .
\nonumber\\
\Sigma[\gamma_{\perp}\to (\gamma_{\parallel})_1 
(\gamma_{\perp})_2]
&=&\omega_1\left(
n_{\parallel}(0)-n_{\perp}(0)\right) < 0\, ,
\eea
where
\bea
n^{''}_{\perp}(0)\equiv \frac{\partial^2}
{\partial \omega^2}n_{\perp}(\omega)|_{\omega=0}\, \simeq\, 
-\frac{1}{m^2}\left(\frac{{\rm B}}{{\rm \tilde{B}}_{\rm cr}}\right)^4\, < 0\, .
\eea
By requiring that $\Sigma\ge 0$, we see that 
the kinematical selection rule allows only for the polarized
reaction $\gamma_{\perp}\to \gamma_{\perp} \gamma_{\perp}$.
We would like to stress that in QED the reaction 
$\gamma_{\perp}\to \gamma_{\perp} \gamma_{\perp}$ is not allowed 
since  $n^{''}_{\perp}(0) >0$, while in
our case this is possible due to the fact that 
$n^{''}_{\perp}(0) < 0$.

Now we will provide an estimation of the absorption coefficient for the 
process $\gamma_{\perp}\to \gamma_{\perp} \gamma_{\perp}$, 
by retaining the leading contributions in $\Delta/m^4\ll 1$ expansion.
The square modulus of the amplitude is obtained by evaluating
the diagram in Eq.(\ref{Mgg}) with 
scalar vertex interaction as in Eq.(\ref{Vertex}). By 
substituting $Z_{+}\to Z_{0}$ and $k^2=M_0^2$ in  Eq.(\ref{M2})
one gets
\bea
|M(\gamma_{\perp}\to \gamma_{\perp} \gamma_{\perp})|^2=
\left(\frac{\omega}{m}\right)^2
\left(\frac{{\rm B}}{{\rm \tilde{B}}_{\rm cr}}\right)^2
\frac{1}{4 \Lambda^2}\left[
(k_1\cdot k_2)(\vec{\epsilon}_1\cdot \vec{\epsilon}_2)
-(\vec{k}_1\cdot \vec{\epsilon_2})(\vec{k}_2\cdot \vec{\epsilon}_1)
\right]^2\, ,
\label{MSQ}
\eea
where higher order terms in $\Delta/m^4$ are neglected. In particular,
we substituted the
denominator in the right-hand-side of the Eq.(\ref{dg2}) with the
leading contribution given by $1/m^4$,
and the wave function renormalization with $Z_0\simeq 1$.
By taking into account the dispersive effects, and the fact that
$\vec{\varepsilon}_{1\perp}\cdot \vec{\varepsilon}_{2\perp}\simeq 1$ 
due to the small opening angle between $\vec{k}_1$ and $\vec{k}_2$, one has
\bea
(k_1\cdot k_2)(\vec{\varepsilon}_{1\perp}
\cdot \vec{\varepsilon}_{2\perp})\, &\simeq&\, 
-\left(\frac{{\rm B}}{{\rm \tilde{B}}_{\rm cr}}\right)^2 \omega_1\omega_2
\nonumber\\
(\vec{k}_1\cdot \vec{\varepsilon}_{2\perp})(\vec{k}_2\cdot 
\vec{\varepsilon}_{1\perp})
&\simeq&\, -\omega_1\omega_2 \sin^2{\theta_{12}}
\,\, \simeq \,\, 
-3\frac{\omega^2}{m^2}
\left(\frac{{\rm B}}{{\rm \tilde{B}}_{\rm cr}}\right)^4 \omega_1\omega_2\, .
\label{terms}
\eea
In the case of
weak magnetic fields ${\rm B}\ll {\rm \tilde{B}}_{\rm cr}$ 
and/or small energies
$\omega\ll m$, one can see from Eq.(\ref{terms}) 
that the second term in parenthesis in Eq.(\ref{MSQ}) is sub-leading with 
respect to the first one and so it can be safely neglected.
Finally, by using Eqs.(\ref{ddg}), (\ref{ph_space}), (\ref{MSQ}), 
(\ref{terms}) and 
integrating over all the phase space, we find the following 
result for the absorption coefficient $d_{\gamma}^{(0)}$ induced
by the massless mode
\bea
d_{\gamma}^{(0)}\simeq \frac{1}{32 \pi}
\left(\frac{{\rm B}}{{\rm \tilde{B}}_{\rm cr}}\right)^6\frac{1}{4\mu^4}
\int_0^{\omega}d\omega_1
\int_0^{\omega}d\omega_2
\left(\omega_1\omega_2\right)^2\delta(\omega-\omega_1-\omega_2)\, .
\eea
Performing the last integral in the phase space and eliminating
the delta-function, we get
\bea
d_{\gamma}^{(0)}=\left(\frac{{\rm B}}{\rm \tilde{B}}_{\rm cr}\right)^6
\left(\frac{\omega}{\mu}\right)^5\left(\frac{\mu}{3840 \pi}\right)\, ,
\label{dg_zero}
\eea
where ${\rm \tilde{B}}_{\rm cr}=\mu^2 $ and $\mu=\sqrt{m\Lambda}$.
The above result holds only provided that 
\bea
\left(\frac{{\rm B}}{\rm \tilde{B}}_{\rm cr}\right)^2
\left(\frac{\omega}{m}\right)^2
\ll 1\, .
\eea

Regarding the photon splitting in the case of 
a scalar coupling with an external electric field, the same results
can be easily obtained from the above expression. At the leading order
in the $\Delta/m^4$ expansion, the expression for the
absorption coefficient is the same as in Eq.(\ref{dg_zero}), 
provided that the external magnetic field is replaced with the electric one,
and the allowed polarizations are
$\gamma_{\parallel}\to \gamma_{\parallel} \, \, \gamma_{\parallel}$ . 

Now we consider the case of a pseudoscalar coupling. If we neglect 
for the moment the effects of the small angles in the final momenta, 
it is not difficult to see that due to the parity violating coupling, the
polarizations of the final photon states will be mainly 
opposite, in particular 
$\gamma_{\parallel}\to \gamma_{\parallel}^{(1)} \gamma_{\perp}^{(2)}$ or 
$\gamma_{\parallel} \to \gamma_{\perp}^{(1)} \gamma_{\parallel}^{(2)}$.
In this case, by taking into account the dispersion effects in the refraction
indices and kinematics, we see that the condition in Eq.(\ref{kin_cond}) 
cannot be satisfied, therefore
the corresponding photon splitting is not allowed in this case.
On the other hand, the process 
$\gamma_{\parallel} \to \gamma_{\parallel}^{(1)} \gamma_{\parallel}^{(2)}$,
due to the pseudoscalar coupling, is forbidden.
Same results hold for the case of an external electric field.

It is interesting to compare the result of the absorption coefficient 
induced by scalar interactions in Eq.(\ref{dg_zero}) 
with the corresponding one due to vacuum polarization effects in QED.
In this last case one has \cite{adler}
\bea
d_{\gamma}^{\rm QED}\simeq 0.1\, 
\left(\frac{{\rm B}}{\rm B^{\rm QED}_{\rm cr}}\right)^6
\left(\frac{\omega}{m_e}\right)^5 {\rm cm}^{-1}\, ,
\eea
where  ${\rm B}^{\rm QED}_{\rm cr}=m_e^2/e =4.41\times 10^{9}$Tesla, 
with $m_e$ and $e$ the electron mass and charge respectively.
If one considers as an example the values 
of scalar mass and coupling around
$m=10^ {-3}$eV and $\Lambda=10^{6}$GeV, as for instance suggested by the
central values of PVLAS data \cite{PVLAS}, 
$\mu =1$MeV, which is incidentally 
quite close to the scale of the critical magnetic field
in QED, namely $\sqrt{{\rm B}_{\rm cr}^{\rm QED}}$.
Then, by using Eq.(\ref{dg_zero}), one gets
\bea
\frac{d_{\gamma}^{(0)}}{d_{\gamma}^{\rm QED}}\simeq 2.8\times 10^{10}\left(
\frac{m_e}{\mu}\right)^{16}\, .
\label{ratio1}
\eea
From Eq.(\ref{ratio1}) follows that 
the QED contribution to the absorption coefficient is smaller than the
scalar/pseudoscalar one, provided that $\mu \lsim 5\times m_e\,$.
We stress that 
this upper bound is stronger than the one obtained by requiring
$(n_{\perp}-1)^{\rm QED}/(n_{\perp}-1)<1$, 
implying that $\mu \lsim 24\times m_e\,$.
However, for the value $\mu=1$MeV, 
as suggested by PVLAS data, the ratio in Eq.(\ref{ratio1}) is 
$\frac{d_{\gamma}^{(0)}}{d_{\gamma}^{\rm QED}} \simeq 4\times 10^5 $
which shows that the massless mode induces quite a large effect
in the photon splitting with respect to the QED one.
It is clear that for laboratory 
magnetic fields of the order of 1 Tesla and
${\rm B}\ll {\rm \tilde{B}}_{\rm cr}$ also the scalar/pseudoscalar 
contribution to 
the absorption coefficient induced by the massless mode 
is quite small, being suppressed by six powers of the external field. 
However, although strongly suppressed,
this effect could have interesting application in 
astrophysical context, where for instance the intensity of magnetic 
field in the core of a supernova could be very large.

The photon splitting through the massive mode remains 
the main mechanism for laboratory experiments, provided the critical condition 
$\xi <1$ for external fields is satisfied.
Indeed, by inspecting the ratio between the absorption coefficients
induced by the massless $(d^{(0)}_{\gamma})$ and massive mode
$(d_{\gamma})$, for external fields far above the critical value 
$\xi \ll 1$, one has
\bea
\frac{d^{(0)}_{\gamma}}{d_{\gamma}}
= \frac{1}{60}\left(\frac{{\rm B}}{\mu^2}\right)^6
\left(\frac{\omega}{m}\right)^6\, .
\label{d_ratio}
\eea
Then, from Eq.(\ref{d_ratio}) it follows that, if
$\mu\gsim \sqrt{\frac{{\rm B}\omega}{2m}}$,
the mechanism of photon splitting by means of 
massive mode is dominant, and, for magnetic fields satisfying the condition
${\rm B } < 50\, m_e^2\, \left(\frac{m}{\omega}\right)$,
it is also larger than the corresponding QED effect. 

Finally, we directly compare the QED background to the photon
splitting process induced by the massive mode, 
when the magnetic fields are far above their critical value $\xi \ll 1$.
In this case one gets,
\bea
\frac{d^{\rm QED}_{\gamma}}{d_{\gamma}}
\simeq 5.4 \times 10^{-62} \left(\frac{\omega}{m}\right)^5
\left(\frac{\mu}{m_e}\right)^4
\left(\frac{\omega}{m_e}\right)
\left(\frac{\rm eV}{m}\right)
\left(\frac{\rm B}{1{\rm Tesla}}\right)^6\, .
\label{d_ratio1}
\eea
In conclusion, 
for laboratory magnetic fields of the
order of Tesla, scalar/pseudoscalar masses and couplings in the range 
$10^{-2} ~{\rm eV} < m  < 10^{2} ~{\rm eV}$ and 
$10^{3} ~{\rm GeV} < \Lambda < 10^{10} ~{\rm GeV}$ respectively, and 
photon beam energy in the range of $1 ~{\rm eV} < \omega  < 10^{2} ~{\rm eV}$, 
the dominant mechanism of photon splitting is through the massive mode, and
the QED background is negligible.

\section{Numerical results}
In this section we present the numerical results of a model independent
analysis based on the photon splitting process in two photons 
mediated by scalar/pseudoscalar interactions. In particular, we will
show that the search for photon splitting process by using
laser experiments in the optical and X-ray frequency range, could allow
to explore extensive regions in the $(m-g)$ plane, depending
on the intensity of the external magnetic field, where 
$g\equiv 1/\Lambda_{S,P}$.
Remarkably, in the case of pseudoscalar (axion) 
coupling, these regions would be complementary to the ones already excluded 
by current laser experiments.

\begin{figure}[tpb]
\begin{center}
\dofig{3.1in}{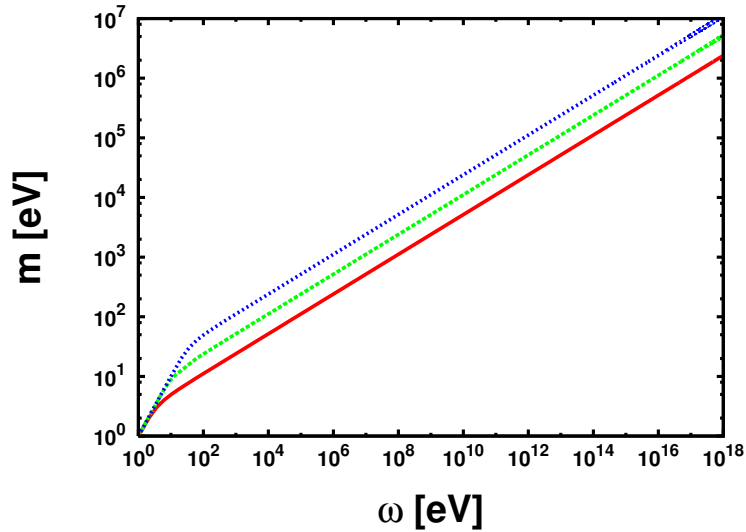}
\end{center}
\caption{\small Upper bounds on $m$ in the scalar case, 
versus the photon energy $\omega$ in eV, for three values 
of external magnetic field $B=0.1,1,10$ Tesla, respectively from bottom to
top.}
\label{Fig3}
\end{figure}

We will consider only the effect of the photon splitting induced by the 
{\it massive} mode, since, as shown in section 4.2, it is the dominant
effect for laboratory experiments.
We emphasize here that,
fixing the values of the photon energy and external field,
the critical conditions in  
Eqs.(\ref{Bcrit_S}), (\ref{Bcrit_P}) set
strong upper bound on the scalar/pseudoscalar masses $m$.
To get a feeling with numbers, in Fig.\ref{Fig3} we show the results for the 
allowed values of scalar masses $m$ versus the photon energy and for
several values of magnetic field.
We consider three representative cases of laboratory 
magnetic fields, namely $|{\rm \vec{B}}|=0.1,1,10$ Tesla.
For example, for photon energies from optical up to the soft
X-ray range, $\omega=$ (1 -200) eV, as for instance the free electron laser
with very high peak brilliance from UV and (soft) X-ray sources 
at DESY and SLAC \cite{ringw2,ringw1},
and magnetic fields between 1-10 Tesla, masses $m$ up to $100$ eV 
can be explored. However, when the photon energy is  
$\omega \gg m $, $\sqrt{|\vec{B}|}$,
the mass upper bound is much below the corresponding $\omega$.

For example, it is possible to realize in laboratory high energy photon beam
by laser back-scattering on a primary electron beam \cite{backscat}, 
as in gamma-gamma colliders \cite{gamma_col}.
However, notice that even with an energy beam of $\omega\simeq 10^5$ GeV,
which is beyond the capability of planned collider experiments,
and a magnetic field of the order of 10 Tesla, one can explore
{\it massive} modes up to $m$ of the order of MeV.
\noindent

Now we will restrict our analysis to the case of 
low energy photons, in particular between the optical and soft 
X-ray range, where the relevant 
scalar/pseudoscalar photon couplings
can only be to photons and neutrinos. For neutrino masses of the order
of meV the $\gamma \to \nu \bar{\nu}$ 
conversion process would be kinematically allowed and the 
$\Gamma$ appearing in Eq.(\ref{dgamma}) 
should be identified with the total width. Then, the formula for the
absorption coefficient of the $\gamma\to \gamma\gamma$ 
process should be multiplied by the corresponding branching ratio.
However, in the case of a PQ axion, the effects of
axion couplings with neutrinos on the total width 
can be neglected due to the fact that 
axion-neutrino coupling is suppressed by neutrino masses.
Indeed, the interaction Lagrangian in this case would be 
$L=g_{A}\bar{\nu}\gamma_5\nu\, \varphi_A$
where $\varphi_A$ is the axion field and $g_{A}\propto m_{\nu}/\Lambda$, 
while $\Lambda$ should be identified with the coupling $\Lambda_P$
appearing in Eq.(\ref{LeffP}).

Since we would like to generalize the results of 
our analysis to the PQ axion case, 
we restrict ourselves to the case where scalar/pseudoscalar particles
are only  effectively coupled to two photons.
The total width will coincide in this case with the width in two photons
given by
\bea
\Gamma=\frac{m^3}{64\pi\Lambda^2}\, .
\label{widthgg}
\eea
The same expression, as a function of mass 
$m$ and coupling $\Lambda$, holds for both scalar and pseudoscalar interactions
in Eqs.(\ref{LeffS}) and  Eqs.(\ref{LeffP}).
The fact the scalar/pseudoscalar particles are only coupled to photons
greatly simplifies the analysis.
In particular we can see that the expression for $\xi$ does not
depend on the coupling $\Lambda$ and it is given by
\bea
\xi=\frac{1}{1024\, \pi^2}\frac{m^{12}}{|{\rm \vec{k}}|^4 |{\rm \vec{B}}|^4
\sin^4{\theta}}
\label{xi1}
\eea
where $|{\rm \vec{k}}|^2\simeq \omega^2-m^2$.
\begin{figure}[tpb]
\begin{center}
\dofigs{3.1in}{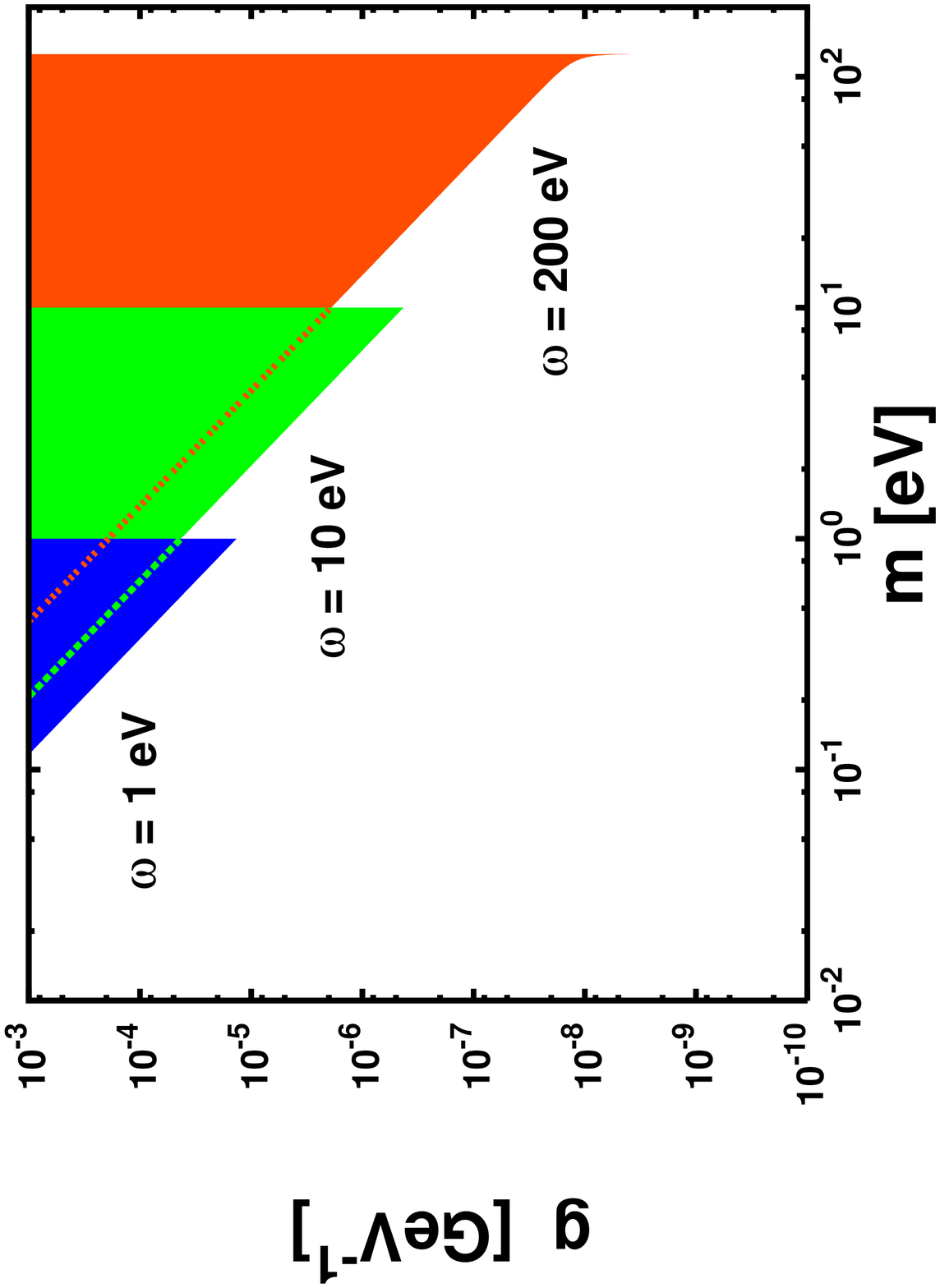}{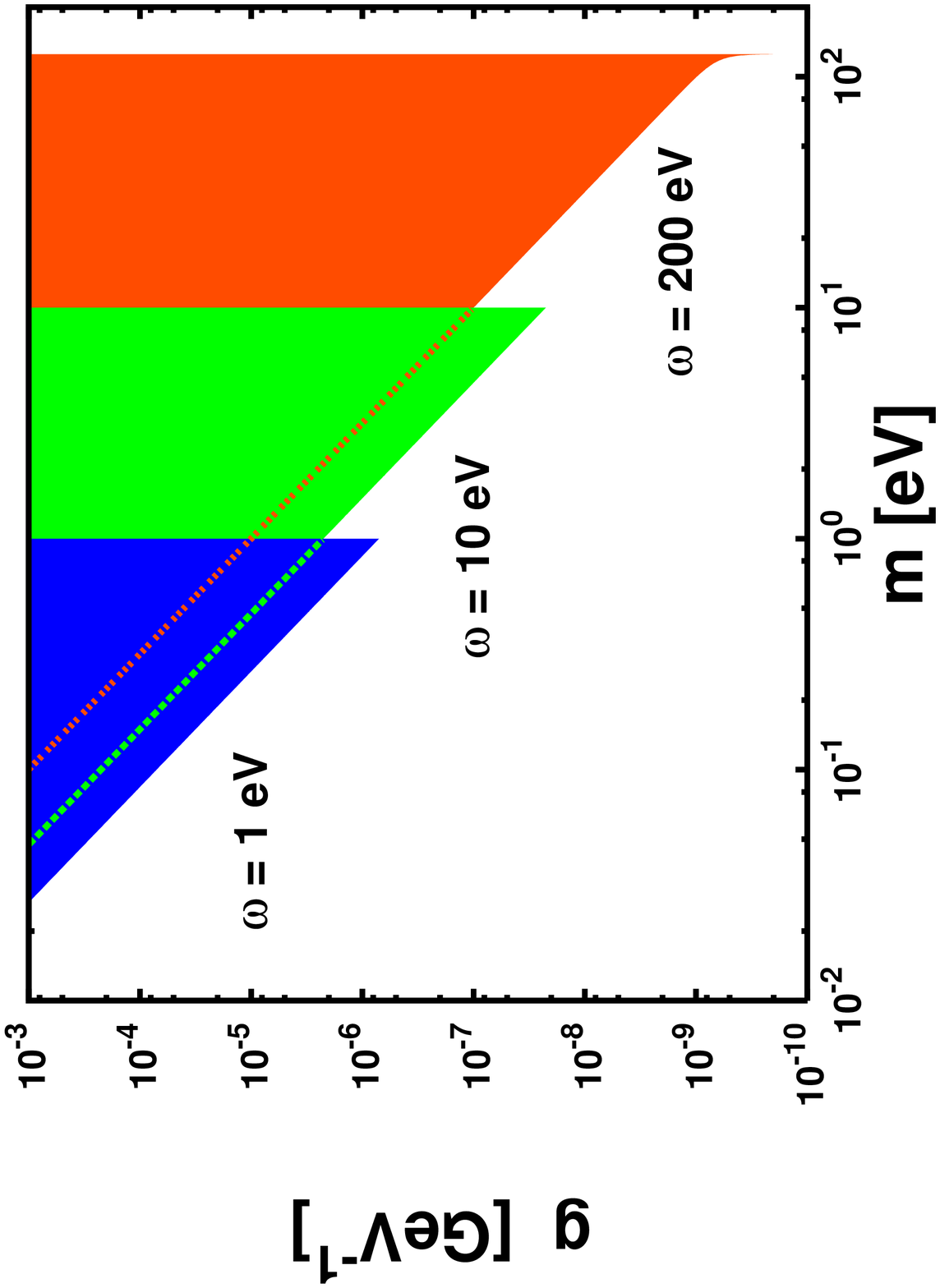}
\end{center}
\caption{\small 
Excluded areas at 95 \% C.L.
in the ${\rm g}=1/\Lambda$ and $m$ plane, corresponding to photon
energies $\omega=1$ eV (blue),   $\omega=10$ eV (green),   
$\omega=200$ eV (red). We have assumed a magnetic field
of 5 Tesla and $L=10$ m, a photon beam with momentum 
orthogonal to the direction of magnetic field 
($\theta=\pi/2$) and number of photons per second
$dN_{\gamma}/dt=10^{18} {\rm s}^{-1}$. 
Results for an integrated running time of 
one day and one year are shown in left and right plots
respectively. }
\label{Fig4}
\end{figure}
\noindent

Now we consider a realistic experiment in which 
a monochromatic photon beam of frequency 
$\omega$ is traveling through a constant and homogeneous magnetic field
of length $L$. If we choose the direction of polarization 
of the magnetic (electric) field component of the photon beam  to be 
parallel to the external magnetic field, $\theta=\pi/2$, 
we select the maximum coupling for 
scalar (pseudoscalar) contributions.
The main signal consists in photon pairs
with different energies $\omega_{1}$ and  $\omega_{2}$,
restricted by the condition of energy conservation $\omega=\omega_1+\omega_2$.
In the case in which $\omega\gg m$, the two photons are produced
both almost forward and parallel to the original momentum of the photon beam, 
with a small opening angle 
$\delta \sim {\cal O}( m/\omega)\ll 1$.\footnote{In detecting the signal, 
one could use a 
particular device where all the photons with energies very close
to the primary photon beam energy $\omega$ can be absorbed. Then, 
by placing a photon detector around the magnet, the split photons
could be detected.
It is beyond the scope of the present paper to 
analyze an efficient way for measuring the photon splitting, and 
this work should be considered as a theoretical proposal.}
Then, excluded regions on the $(m-g)$ plane can be set, for instance,
by requiring that no significant number of events  are observed 
at 95\% confidence level.
Since this process has practically no background, due to the fact that 
in QED this effect is very suppressed, the corresponding significant
number of 
standard deviation associated to a number of observed events $N_S$ would be 
of the order $\sqrt{N_S}$. In the particular case of 95\% C.L. this  
implies $N_S\simeq 4$.
Then, the requirement that no significant number of events are 
observed at 95\% C.L. would imply
\bea
d_{\gamma} < \frac{4}{N_{\gamma} L}
\label{constr}
\eea
where $d_{\gamma}$ is given in Eq.(\ref{dgamma}), with $\Gamma$ and $\xi$
given by Eqs.(\ref{widthgg}) and (\ref{xi1}) respectively.
For $N_{\gamma}$ we have 
assumed a representative laser brilliance corresponding to 
$dN_{\gamma}/dt=10^{18}/{\rm sec}$, 
as for instance in the case of free electron lasers.

Excluded regions in the $(m-g)$ plane corresponding 
to the upper limit in Eq.(\ref{constr}) are
shown in Fig.\ref{Fig4} for a magnetic field of 5 Tesla and for
different values of photon beam energy, namely $\omega=1,~10,~200$ eV.
Here in the left (right) plot we show the exclusion regions 
corresponding to one day (year) of running time, respectively.
As we can see from these results, by increasing the photon energy,
larger values of $\Lambda$  can be probed. In particular,
with laser frequencies in the optical range $\omega=1$ eV, values of
$\Lambda \lsim 10^{6}$ GeV can be explored after one year of 
running, while $\Lambda \lsim 10^{9}$ GeV can be reached 
with $\omega=200$ eV.
The range of masses that can be explored with a 5 Tesla magnetic field,
are just limited by the photon energy, namely 
$m\le \omega$, for $\omega=1,10$ eV.
However, for this value of the magnetic field and 
$\omega=200$ eV, the resonant region ($\xi\simeq 1$) is achieved below
the kinematical upper bound $m\le \omega$, and so for $\omega=200$ eV
only the range of masses up to $m \lsim 125$ eV can be explored.
Around the region close to the critical point, 
the $d_{\gamma}$ becomes insensitive 
to $\Lambda$ as shown in the previous section, see Eq.(\ref{imPmax}).
This is the reason why the shape of the 
lower part of the red area, corresponding to $\omega=200$ eV,
near the end point of $m$ is different from
the other cases. The narrow width characterizing the end point region 
with large values of $\Lambda$, is of the order of $\Gamma$.

\begin{figure}[tpb]
\begin{center}
\dofigspec{3.1in}{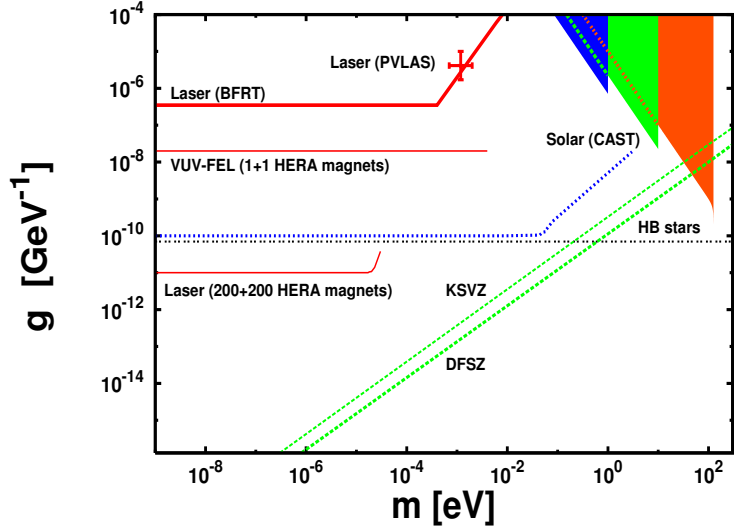}
\end{center}
\caption{\small Results of right plot in Fig.\ref{Fig4} embedded 
in the ${\rm g}=1/\Lambda$ and $m$ plane with exclusion regions
coming from laser experiments BFRT \cite{BFRT} and
PVLAS \cite{PVLAS}. Exclusion regions coming from solar experiments
searches for axion from the sun (Solar CAST) 
\cite{SolarCast} and the horizontal branch stars (HB stars) \cite{HBstars}
are also shown.
The inclined lines (green) correspond to the
predictions in DFSZ \cite{dfsz} and KSVZ \cite{ksvz} axion models. 
See text for further details. }
\label{Fig5}
\end{figure}
\noindent

Let us now discuss the consequences of these results for the axion physics.
We remind here that the mass of the axion field
connected to the PQ symmetry is related to the mass and decay
constant of the pion by \cite{WW}
\bea
m_A=z^{1/2}/(1+z)\, m_{\pi}\, f_{\pi}/f_A=0.6\, {\rm meV}\times
(10^{10}{\rm GeV}/f_A)
\eea
where $f_A\gg 247$ GeV is the scale where this symmetry is broken, 
$f_{\pi}$ and $m_{\pi}$ are the decay constant and mass
of the pion and the current quark mass ratio $z=m_u/m_d$.
The couplings of the axion to SM particles are not only
functions of the scale $f_A$, but also model dependent. 
For instance, the constant $f_A$ is connected to the photon coupling 
$g_{A\gamma}=\Lambda_P^{-1}$ in 
Eq.(\ref{LeffP}) by axial anomaly, and it is given by
\bea
g_{A\gamma}=-\frac{\alpha}{2\pi\,f_A}\left(\frac{E}{N}-\frac{2}{3}
\frac{4+z}{1+z}\right)
\eea
where $E/N$ is the ratio of electromagnetic over color anomalies and it
is model dependent. The most popular models are DFSZ \cite{dfsz}, 
where axion is
embedded in Grand Unified Theories, and KSVZ axion is the so called
``hadronic'' class of axions \cite{ksvz}. In both these models the couplings 
$g_{A\gamma}$ are predicted in terms of $f_A$ and are slightly different.
The most stringent constraints on $g_{A\gamma}$ comes from 
cosmological and astrophysical arguments.
However, there is a general class of experiments with laser that aim
to produce axions in laboratory. They can be divided in two classes,
one connected with photon-regeneration and another one that analyzes the
indirect axion effects on light propagation. The first one is based on 
the Primakoff process \cite{Prim}, 
where a photon is converted in external magnetic field
and then axions can be re-converted in photons with same mechanism 
after passing through a wall where all primary photons are stopped.
The latter one probes the axion-coupling by measuring the change in the
polarization of the photon after passing through magnetic field.

In Fig.\ref{Fig5} we report the comparison between the
present 
excluded areas in the $(m-g)$ plane in the case of axion searches, by
embedding the results contained in the right plot of Fig.\ref{Fig4}
corresponding to one year of integrated running time.
The predictions of the models of DFSZ and  KSVZ axions are also shown.
The exclusion regions based on the solar experiments for axions from the Sun 
\cite{SolarCast} (indicated with Solar (CAST) = CERN Axion Solar Telescope) 
and the constraints from 
horizontal branch (HB) stars \cite{HBstars} arising from a consideration of 
stellar energy losses through axion production are reported.
Recently, the PVLAS experiment \cite{PVLAS}, 
reported the observation of a rotation of 
the polarization plane of light propagating through a transverse 
static magnetic field. This is shown as a point with error bars on 
the border of laser area in  Fig.\ref{Fig5}.
Notice that the signal seen in \cite{PVLAS} 
is incompatible with the exclusion area from 
experiment \cite{BFRT}.
Constraints coming from the galactic dark matter experiments \cite{galactic}
are not reported in Fig.\ref{Fig5}. 
However, this class of experiments 
strongly constrain only a narrow region in the 
$(m-g)$ plane corresponding to 
axion masses in the range $10^{-6} {\rm eV} \lsim m_a \lsim 10^ {-5} {\rm eV}$.
We have also included the results of a recent proposal \cite{ringw2,ringw1}
of  laser experiment based on high-energy photon regeneration mechanisms, 
which is the X-ray analogous of the optical one.
The free-electron lasers (FEL) at DESY's TESLA and dipole magnets of the
type used in DESY's electron-proton collider HERA have been employed.
The two horizontal lines correspond to the experimental analysis based on 
1+1 and 200+200 HERA dipole magnets.

As clear from the plot, the new areas in the $(m-g)$ plane
that can be probed by searching 
for photon splitting in two photons would allow to cover
complementary areas not achievable by other two classes of 
laser experiments. It is worth noticing that both the predictions of
KSVZ and DFSZ models fall inside 
the area that could be probed by the $\omega=200$ eV laser experiment 
proposed here.
As we can see from these results, the astrophysical constraints rule out
a large region of the parameter space including the one 
we are interested in, as well as the region covered by laser experiments.
However, recently it has been shown that in certain models these
astrophysical constraints can be evaded \cite{astro_evas}.

Finally, we would like to emphasize that in the case in which a 
significant number of events 
of photon splitting $\gamma\to \gamma\gamma$ should be found, the values 
of the scalar/pseudoscalar mass and width can be easily reconstructed.
This would allow to precisely tune the external magnetic field and 
photon energy in order to approach the resonant region, where the
effect of photon splitting could be largely amplified.
Regarding the origin of the interaction, if
there is a scalar or pseudoscalar coupling, this can be disentangled
by analyzing the polarization 
of the final photons with respect to the direction of the magnetic field
or by analyzing their angular distributions, 
provided that a sufficient number of events are observed.

Let us also shortly discuss the case of the Higgs boson effects in 
this context. We start with the case of high energy photon 
beam traveling in a constant magnetic field and with the coupling to a 
light Higgs boson of mass $m_H\simeq 120$ GeV. The main decay mode
of the Higgs in this scenario is in $b\bar{b}$.
In the effective propagator there would be a very light mode, 
and a {\it massive} mode of the size of the Higgs mass.
In order to generate the $\gamma\to b\bar{b}$ decay in external field
one needs first to check if the condition (\ref{cond1}) is satisfied
for the corresponding {\it massive} mode.
As we discussed in the previous sections this would require too high
critical magnetic fields, when the energy of the 
photon beam is of the order of TeV,
which cannot be realized in 
the laboratory. Nevertheless, there might be the possibility of 
a sizeable effect on the photon splitting 
mediated by the Higgs effective coupling
in the framework of astrophysics, for example in the core of
a supernova, where magnetic fields are very large. However, in that 
case the approximation of considering only  constant and
homogeneous magnetic fields is not correct, since also plasma 
effects should be taken into account. 
When magnetic fields are inhomogeneous, momentum can be absorbed 
and the Higgs boson could in principle be produced on-shell by 
means of the Primakoff process.
Although the analysis of this latter effect 
in astrophysical context should be quite 
interesting, it goes beyond the scope of the present paper.

\section{Conclusions}

In this paper we have investigated the analytical properties of the
effective photon propagator in an external homogeneous and static 
magnetic or electric fields, in the presence of scalar/pseudoscalar
couplings. These results have 
been obtained by summing up in the photon propagator the relevant class of 
Feynman diagrams.  They include the ones with 
mixing term of photon with scalar/pseudoscalar 
fields, where in the scalar/pseudoscalar propagator
the corresponding self-energies have been exactly summed up.
Then, we analyzed the solutions of
the associated dispersion relations.

Due to the presence of an external field,  the standard
photon dispersion relations in vacuum will be modified.
While the effect of the real part of scalar/pseudoscalar 
self-energy can be re-absorbed in the corresponding mass term of
the scalar/pseudoscalar, the imaginary 
part, connected to the corresponding width $\Gamma$, could play a crucial 
role in the photon dispersion relations giving rise to a
non-vanishing contribution to the imaginary part of the effective 
photon propagator. In other words, the presence of $\Gamma$ can induce 
a non-vanishing contribution for photon absorption coefficient.

As known, two new propagation modes are allowed for the photon, 
in addition to the usual one which is not affected by the external field
contributions.
The lightest one is associated to a light-like
mode with refraction index $n> 1$, while the heaviest one corresponds
to a {\it massive} mode with 
mass of the order of the scalar/pseudoscalar one.

The new aspect of our work with respect to previous studies 
concerns the inclusion of the effects 
of the imaginary part of scalar/pseudoscalar
self-energy in the effective photon propagator. 
In particular, we have shown that the {\it massive} mode 
can be allowed only when the
contributions of the scalar/pseudoscalar width $\Gamma$ 
are smaller or comparable 
to the mixing effects, see Eq.(\ref{cond1}).
This will give rise to a critical condition 
for the external field, which is absent in the limit $\Gamma\to 0$,
depending on the photon energy, scalar/pseudoscalar
masses and the two-photon coupling.
A finite {\it width} for the photon can then be induced for
the {\it massive} mode, being proportional to $\Gamma$,
generating a non-vanishing value for the photon absorption coefficient
$d_{\gamma}$.

Although very small, being suppressed by the scalar/pseudoscalar width,
the $d_{\gamma}$ can be sizeably enhanced due to a 
strong resonant phenomenon. In particular, we have found that 
a potentially large contribution to the
splitting conversion $\gamma\to \gamma\gamma$ could be generated from the
scalar/pseudoscalar decay in two photons when the external field
approaches its critical value.

We have also analyzed the contribution to the absorption coefficient induced 
by the massless photon mode. In the case of an external magnetic field
and a scalar interaction, 
we find that the probability of photon splitting in two photons 
turn out to be very small, being 
suppressed by $({\rm B}/{\rm B_{cr}})^6$ where (in relativistic
unities) ${\rm B_{cr}}=m\Lambda$.
Nevertheless, this effect, depending on mass $m$ and coupling $\Lambda$,
could be much larger than the one induced by the 
QED vacuum polarization in the presence of an external magnetic field.
For instance, for  characteristic 
values of $m=10^{-3}$eV and $\Lambda=10^6$GeV, we find that $d_{\gamma}$
is about five order of magnitude larger than the corresponding QED one.
The leading contribution to $d_{\gamma}$ is provided by the 
small effects of dispersions in the refraction index.
For the case of a pseudoscalar interaction, due to the parity violating
coupling and kinematic factors, the photon splitting induced by the massless
mode is forbidden.

We have analyzed the consequences of these results for 
a new kind of laser laboratory experiments, based on the searching 
of the photon splitting 
$\gamma\to \gamma\gamma$ conversion in external constant and homogeneous 
magnetic field. In particular, we have considered the photon splitting
phenomenon induced by the massive mode.
By taking the case of high brilliance lasers with
$dN_{\gamma}/dt=10^{18}$/sec, in the range between optical 
($\omega \simeq 1\, {\rm eV}$) 
and the low X-ray  ($\omega\simeq 200$ eV) frequencies, magnetic
fields of 10 meters long and of the order of 5 Tesla, 
we show that it is possible to probe a large area of the 
$m$ and $g=1/\Lambda$ plane, provided that the process
$\gamma\to\gamma\gamma$ can be efficiently detected. The probed areas 
for scalar and pseudoscalar case are slightly different.
We find that in the
case of  $\omega\simeq 200$ eV,
scalar/pseudoscalar masses can be probed up to $m<100$ eV , while
in the case of  $\omega=1$ eV the region up to the maximum value 
allowed by the energy conservation, i.e.  $m< 1$ eV can be explored.
Moreover, the sensitivity on the scale $\Lambda$ associated to the 
two-photon coupling can become , in one year of running time,
close to $\Lambda\simeq 10^{6}$ GeV and $\Lambda\simeq 10^{9}$ GeV, 
corresponding to the case of $\omega=1$ eV 
and  $\omega=200$ eV  laser frequencies, respectively.

We have also compared our results with the present bounds from
axion searches.
In particular, by restricting our predictions 
to the case of pseudoscalar couplings, 
we found that a large area in the $m$ and $g$ plane
of the axion can be tested. 
Remarkably, this area is complementary to the ones
already explored by the present laser experiments, like PVLAS and BFRT,
and to the sensitivity area of new 
recent proposal based on the photon-regeneration class of experiments
with X-ray free electron laser facility at HERA.
We have also shown that some predictions of the KSVZ and DFSZ models 
could be tested by means of $\gamma\to \gamma\gamma$ searches in 
external magnetic fields.

\vspace{1cm}
{\Large\bf Acknowledgments}\\
We would like to acknowledge useful discussions with D. Anselmi,
M. Chaichian, M. Giovannini, G.F. Giudice, 
K. Kajantie, B. Mele, P.B. Pal, R. Rattazzi, P. Sikivie, L. Stodolsky, and
K. Zioutas.
E.G. would like to thank the CERN TH-division for kind
hospitality during the preparation of this work. S.R. thanks the
ASICTP for kind hospitality during the final stages of this work.
This work is supported by the Academy of Finland (Project number 104368).

\vspace{1cm}
\section*{Appendix A}
In this appendix we report the exact solutions of the {\it mass-gap} equation
(\ref{master2}). This is a cubic algebraic equation and can be analytically
solved by using the known cubic formula. In particular, 
by substituting $k^2=m^2+\varepsilon$ in  Eq. (\ref{master2}), this can 
be simplified as 
\bea
x^3+x^2+x\left(\xi_{\Gamma}-\xi_{\Delta}\right)+\xi_{\Gamma}=0\, ,
\eea
where $x=\varepsilon/m^2$,
$\xi_{\Gamma}=\Gamma^2/m^2$ and $\xi_{\Delta}=\Delta/m^4$.
Notice that in writing the gap equation 
we have identified the ${\rm Im} \Sigma(k^2)$ with 
$-m\Gamma$. In general this statement is not correct since 
the imaginary part of 
scalar/pseudoscalar width depends on $k^2$, in particular
for the photon splitting it is proportional to $(k^2)^2/\Lambda^2$,
see Appendix B for more details.
However, the approximation to set ${\rm Im} \Sigma(k^2)=-m\Gamma$  
is valid only
when $k^2\sim {\cal O}(m^2)$, that is for the massive solutions. 
On the other hand, for the 
massless mode where, $k^2\sim {\cal O}(\Delta/m^2)$, the
${\rm Im}\Sigma(k^2)$ can be neglected with respect to the $m^2$ term
and one can safely switch off the ${\rm Im}\Sigma(k^2)$ in the
scalar/pseudoscalar propagator provided that $\Delta/m^4\ll 1$.
Therefore, in order to simplify the problem, one can 
solve the gap equation by using 
${\rm Im} \Sigma(k^2)=m\Gamma$ and then, 
in order to recover the correct result for the massless mode,  set 
$\xi_{\Gamma} \to 0$ on the corresponding massless solution.

By using this approach, we solve the cubic equation above and 
find the following expressions for the solutions of  Eq. (\ref{master2}),
namely $k^2=M^2_0$ and  $k^2=M^2_{\pm}$ with
\bea
M^2_0&=&\lim_{\xi_{\Gamma}\to 0}\,  m^2\left(\frac{2}{3}+X_{+}+X_{-}\right) 
\nonumber\\
M^2_{\pm}&=&
m^2\left(\frac{2}{3}-\frac{1}{2}\left(X_{+}+X_{-}\right)
\pm i\frac{\sqrt{3}}{2}\left(X_{+}-X_{-}\right)\right)\, ,
\label{exact}
\eea
where
\bea
X_{\pm}=\left(R\pm\sqrt{D}\right)^{1/3},~~~~~~
D=Q^3+R^2
\eea
and in our case
\bea
Q=\frac{3\left(\xi_{\Gamma}-\xi_{\Delta}\right)-1}{9},~~~~~~~
R=-\frac{9\left(2\xi_{\Gamma}+\xi_{\Delta}\right)+2}{54}\, .
\eea
The massive solutions $k^2=M_{\pm}^2$ are real only 
when the condition $D<0$ is satisfied, where
\bea
D=\frac{1}{108}\left\{-\xi_{\Delta}^2+4\, \xi_{\Gamma}-4\,\xi_{\Delta}^3
+20\,\xi_{\Delta}\,\xi_{\Gamma}+12\,\left(\xi_{\Delta}^2\,\xi_{\Gamma}
-\xi_{\Gamma}^2\,\xi_{\Delta}\right)
+8\,\xi_{\Gamma}^2+4\,\xi_{\Gamma}^3\right\}\,.
\label{D}
\eea
In our problem, the parameters $\xi_{\Delta}\ll 1$ and $\xi_{\Gamma}\ll 1$ are 
very small quantities. We remind here that these are
proportional to inverse powers of the effective coupling
$\Lambda$ with two photons, 
which is assumed to be much larger than the typical
scales appearing in the numerator of $\xi_{\Gamma}$ and  $\xi_{\Delta}$,
namely the photon energy $\omega$, momentum $\vec{k}$, and external field.
From a first inspection of Eq.(\ref{D}), one can see that, 
in order to have real solutions ($D<0$), these parameters should
satisfy the following hierarchy
$\xi_{\Gamma}\sim \xi_{\Delta}^2$.
This is also consistent with the fact that $\xi_{\Gamma}$ 
and $\xi_{\Delta}$ scale as
$\xi_{\Gamma}\sim 1/\Lambda^4$ and 
$\xi_{\Delta} \sim 1/\Lambda^2$ respectively as a function of $\Lambda$.
In particular, by dropping the 
terms of order ${\cal O}(\xi_{\Delta}^3)$ in Eq.(\ref{D}), the condition
$D<0$ is equivalent to 
\bea
\xi_{\Delta} > \xi_{\Delta}^{\rm c\,  (0)}\equiv  2\sqrt{\xi_{\Gamma}}\, 
\label{crit}
\eea
which is exactly
the inequality in (\ref{cond1}) derived under the approximation of
neglecting terms of order $\xi_{\Delta}^3$.

There is only one solution of  $D=0$, corresponding to 
real and positive values of $\xi_{\Delta}$ and $\xi_{\Gamma}$, and it is
given by
\bea
\xi_{\Gamma}&=&\xi_{\Delta}-\frac{2}{3}+\frac{2\left(1-27
\xi_{\Delta}\right)}{3H(\xi_{\Delta})}+\frac{H(\xi_{\Delta})}{6}\, ,
\label{solGap}
\eea
where the expression for the function $H(x)$ is
\bea
H(x)&=&\left(8+540x-729x^2+
3\,\sqrt{3}\,\sqrt{x}\left(8+27x\right)^{3/2}
\right)^{1/3}\, .
\eea
As we can see from these results, there always exists a critical value
for the external fields for which the {\it massive} 
solutions become degenerate, implying
the appearance of a double pole in the effective photon propagator.

Now we will provide below the corresponding solutions obtained by expanding the
exact ones in terms of the small parameters $\xi_{\Delta}$ and 
$\xi_{\Gamma}$.
Let us first
define $\xi_{\Delta}^{\rm c}$ being the value of $\xi_{\Delta}$
satisfying the exact solution of Eq.(\ref{solGap}).
Then, by expanding $\xi_{\Delta}$ 
in powers of $\delta=\Gamma/m$, we obtain, up to terms
${\cal O}(\delta^{6})$, the following expression
\bea
\xi_{\Delta}^{\rm c}=\xi_{\Delta}^{\rm c\, (0)}\, 
\Big(1+R(\delta)\,\Big)
\label{crit1}
\eea
where
\bea
R(x)=x-
\frac{1}{2}\,x^2+x^{3}-\frac{5}{2}\,x^{4}
+8\,x^{5}+ {\cal O}(x^{6})\, .
\eea

Finally, the corresponding next-to-leading order corrections,
in $\xi_{\Delta}$ expansion, to the solutions 
in Eqs.(\ref{gap_sol}) are given by
\bea
M_0&=&-\frac{\Delta}{m^2}\left(1-\xi_{\Delta}+2\xi_{\Delta}^2\right)
\nonumber\\
M^2_{\pm}&=&m^2\left\{1+\frac{\xi_{\Delta}}{2}\left(1\pm\sqrt{1-\xi}\right)
-\frac{\xi_{\Delta}^2}{2}\left(1\pm\frac{2-\xi}{2\sqrt{1-\xi}}\right)\right\}
\label{NLOsol}
\eea
where the symbol $\xi=4\xi_{\Gamma}/\xi_{\Delta}^2$ 
is the same as appearing in Eq.(\ref{xidef}). Notice that for 
the solutions in (\ref{NLOsol}) 
the range $0<\xi <1$ is not valid anymore, 
and, according to Eq.(\ref{crit1}), must be modified as 
$0\le \xi \le 1-\delta +{\cal O}(\delta^2) $.
It can be easily checked that 
for the maximum value $\xi^{\rm max}=1-\delta$,
the difference $M^2_{+}-M^2_{-}$ vanishes up to terms 
of order ${\cal O}(\xi_{\Delta}^3)$, within the validity of 
the approximated solutions.

Regarding the massless mode, the expression for $M_0^2$ can be 
easily re-obtained by setting to zero the scalar/pseudoscalar
width $\Gamma$ directly in the gap equation. 
This will reduce the cubic equation in Eq.(\ref{master2}) 
to a second order one.
In particular, by setting  ${\rm Im}\Sigma(k^2)$ to zero in 
Eq.(\ref{master2}), one gets a more compact solution for the massless mode,
equivalent to the first solution in Eq.(\ref{exact}), given by
\bea
M_0^2=\frac{m^2}{2}\left(1-\sqrt{1+4\xi_{\Delta}}\right)\, ,
\label{M0_exact}
\eea
where the leading orders in $\xi_{\Delta}$ expansion easily recover the
corresponding results in Eqs.(\ref{gap_sol}) and (\ref{NLOsol}).
\section*{Appendix B}
In this appendix we provide the calculation at 1-loop 
of the imaginary part of scalar (pseudoscalar) self-energy $\Sigma_{S(P)}$, 
induced by the interactions in Eqs.(\ref{LeffS}),(\ref{LeffP}).
While the real part of $\Sigma_{S(P)}$ is divergent, its 
imaginary part is finite at 1-loop and can be easily calculated 
by making use of the optical theorem. 
In particular, the ${\rm Im}[\Sigma_{S(P)}(k^2)]$ is connected to the
square modulus of the off-shell scalar/pseudoscalar decay in two on-shell 
photons, which schematically can be represented as
\bea
\epsfbox{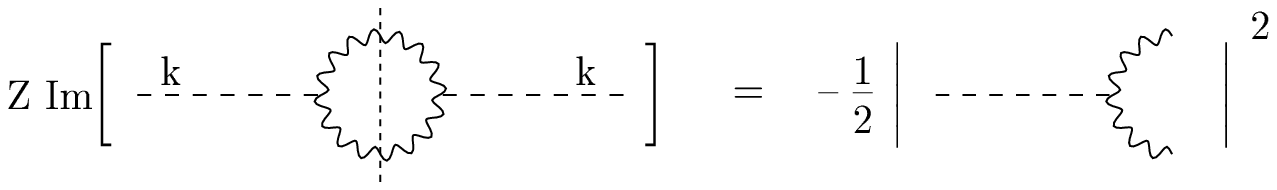}
\label{Optical}
\eea
where $Z$ is the renormalization constant of 
scalar/pseudoscalar field,
the continuous and dashed line stands for the scalar/pseudoscalar 
and photon fields respectively, and $k,k_{1,2}$ are the corresponding
momenta.
In the left-hand side of (\ref{Optical}), 
the vertical dashed lines correspond to the Cutkosky rule which 
selects the imaginary part of the photon propagators in the loop, by
replacing in each of them  $1/(p^2+i\epsilon)\to 
-\pi \delta(p^2) \theta(\omega)$, with 
$p^2\equiv p_{\mu}p^{\mu}$ and $\omega\equiv p_0$,
where $p_{\mu}$ 
indicate the generic 4-momenta of photon lines in the loop.
The factor $1/2$ in front of the amplitude 
square is due to symmetry factor connected to the two photon lines.
By using the Cutkosky rule and the fact that at this order 
in perturbation theory $Z=1$, 
one obtains
\bea
{\rm Im}[\Sigma_{S(P)}(k^2)] &=&-\frac{1}{2}
\frac{d^3 k_1}{(2\pi)^3 2\omega_1}
\frac{d^3 k_2}{(2\pi)^3 2\omega_2} (2\pi)^4\delta^4(k-k_1-k_2)\,
\sum_{\rm pol} \, \left| 
V_{\gamma\gamma}^{S(P)}(k_1,k_2)\right|^2
\label{ImSigma}
\eea
where $V_{\gamma\gamma}^{S(P)}(k_1,k_2)$ is 
the two photon vertex for the scalar(pseudoscalar) field given in 
Eq.(\ref{Vertex}), 
and the $k_{1,2}$ are the corresponding final photon momenta on shell,
$k_{1,2}^2=0$. The sum $ \sum_{\rm pol}$
is extended to all photon polarizations.
Notice that, as an approximation, 
we have not considered the effect of the external field
corrections, and so the photon dispersion relations are not modified. 

Finally, by using the results in Eq.(\ref{Vertex}), we get, for 
the contribution at 1-loop to the self-energy,
\bea
{\rm Im}[\Sigma_{S(P)}(k^2)]=-\frac{(k^2)^2}{64\pi\Lambda^2_{S(P)}}\, .
\label{ImSigma2}
\eea
When Eq.(\ref{ImSigma}) is evaluated on-shell the 
scalar(pseudoscalar) width $\Gamma_{S(P)}$ on-shell is recovered
by means of the relation ${\rm Im}[\Sigma(k^2)]|_{k^2=m^2}=-m\Gamma$, in
particular
\bea
\Gamma_{S(P)}=\frac{m^3}{64\pi\Lambda^2_{S(P)}}\, ,
\eea
which is in agreement with  Eq.(\ref{widthgg}).

\end{document}